\documentclass[fleqn,usenatbib]{mnras}

\usepackage{newtxtext,newtxmath}

\usepackage[T1]{fontenc}
\usepackage{ae,aecompl}

\usepackage{graphicx}	
\usepackage{amsmath}	

\title[Transmission spectrum of WASP-88b]{Transmission spectroscopy with VLT FORS2: a featureless spectrum for the low-density transiting exoplanet WASP-88b}

\author[P. Spyratos et al.]{Petros Spyratos$^{1}$\thanks{E-mail: p.spyratos@keele.ac.uk},
Nikolay Nikolov$^{2}$,
John Southworth$^{1}$,
Savvas Constantinou$^{3}$,
\newauthor
Nikku Madhusudhan$^{3}$,
Aarynn L. Carter$^{4}$,
Ernst J. W. de Mooij$^{5}$,
Jonathan J. Fortney$^{4}$,
\newauthor
Neale P. Gibson$^{6}$,
Jayesh M. Goyal$^{7}$,
Christiane Helling$^{8}$,
Nathan J. Mayne$^{9}$ and
\newauthor
Thomas Mikal-Evans$^{10}$
\\
$^{1}$Astrophysics Group, Keele University, Staffordshire, ST5 5BG, UK\\
$^{2}$Space Telescope Science Institute, 3700 San Martin Dr, Baltimore, MD 21218, USA\\
$^{3}$Institute of Astronomy, University of Cambridge, Madingley Road, Cambridge CB3 0HA, UK\\
$^{4}$Department of Astronomy and Astrophysics, University of California, Santa Cruz, 1156 High St, Santa Cruz, CA 95064, USA\\
$^{5}$Astrophysics Research Centre, School of Mathematics and Physics, Queens University Belfast, Belfast BT7 1NN, UK\\
$^{6}$School of Physics, Trinity College Dublin, Dublin 2, Ireland\\
$^{7}$Department of Astronomy and Carl Sagan Institute, Cornell University, 122 Sciences Drive, Ithaca, NY 14853, USA\\
$^{8}$Centre for Exoplanet Science, University of St Andrews, Nort Haugh, St Andrews, KY169SS, UK\\
$^{9}$Astrophysics Group, University of Exeter, Exeter, EX4 2QL, UK\\
$^{10}$Kavli Institute for Astrophysics and Space Research, Massachusetts Institute of Technology, Cambridge, MA 02139, USA\\
}

\date{Accepted XXX. Received YYY; in original form ZZZ.}

\pubyear{2021}

\begin{document}
\label{firstpage}
\pagerange{\pageref{firstpage}--\pageref{lastpage}}
\maketitle

\begin{abstract}
We present ground-based optical transmission spectroscopy of the low-density hot Jupiter WASP-88b covering the wavelength range $4413-8333$\,\AA~with the FORS2 spectrograph on the Very Large Telescope. The FORS2 white light curves exhibit a significant time-correlated noise which we model using a Gaussian Process and remove as a wavelength-independent component from the spectroscopic light curves. We analyse complementary photometric observations from the Transiting Exoplanet Survey Satellite and refine the system properties and ephemeris. We find a featureless transmission spectrum with increased absorption towards shorter wavelengths. We perform an atmospheric retrieval analysis with the \textsc{aura} code, finding tentative evidence for haze in the upper atmospheric layers and a lower likelihood for a dense cloud deck. Whilst our retrieval analysis results point toward clouds and hazes, further evidence is needed to definitively reject a clear-sky scenario.
\end{abstract}




\section{Introduction}

Since the first detection of a constituent in the atmosphere of a planet outside the Solar system \citep{2002ApJ...568..377C}, transmission spectroscopy has become the central tool for probing the atmospheric composition and structure  of transiting exoplanets. During a planetary transit, part of the observed starlight filters through the planetary atmosphere depending on the composition. This causes small, wavelength-dependent variations in the apparent planet radius, which can inform us about the physical and chemical conditions of exoplanet atmospheres. Model spectra of irradiated hot Jupiter atmospheres free of clouds predict broad spectral signatures of Na and K in the optical and strong molecular absorbers e.g. H$_2$O in the infrared and, for very hot environments, TiO and/or VO in the optical \citep{2000ApJ...537..916S,sudarsky00,2001ApJ...553.1006B,2003ApJ...594.1011H,2010ApJ...709.1396F}. However, increased opacity from clouds can effectively reduce the strength of absorption features across the entire optical wavelength range and excess scattering from hazes can add an increasing slope with decreasing wavelength. Such mechanisms are often associated with flat transmission spectra \citep[e.g.][]{2013MNRAS.428.3680G,2013MNRAS.436.2974G,2017MNRAS.467.4591G,2019MNRAS.482.2065E,2020MNRAS.497.5155W}. Observational constraints for each scenario are key to our understanding of the diversity of exoplanetary atmospheres, the processes of planetary formation and evolution, as well as the formation and occurrence of clouds and hazes.

The \textit{Hubble} and \textit{Spitzer Space Telescopes} (\textit{HST} \& \textit{Spitzer}) have been paramount in the field with a multitude of atmospheric compositional constraints including absorption from atomic \citep[e.g.][]{2008ApJ...686..658S,2015MNRAS.446.2428S,2014MNRAS.437...46N} and molecular  species \citep[e.g.][]{2013ApJ...774...95D,2013MNRAS.434.3252H,2013MNRAS.435.3481W,2016Natur.529...59S,2019A&A...622A..71V}. Significant progress has also been made from the ground, exploiting techniques such as broadband photometry \citep[e.g.][]{2013A&A...553A..26N,2013MNRAS.436....2M}, long-slit \citep[e.g.][]{2012MNRAS.426.1663S} and multi-object spectroscopy \citep[e.g.][]{2010Natur.468..669B,2013MNRAS.428.3680G,2017MNRAS.467.4591G,2016ApJ...832..191N}. The latter method has been extensively applied to a number of hot Jupiters using the FOcal Reducer Spectrograph \citep[FORS2,][]{1998Msngr..94....1A,2016SPIE.9908E..2BB}, installed on the Very Large Telescope (VLT), with success in placing constraints on the abundances of Na and K and distinguishing clear from cloudy and hazy hot-Jupiter atmospheres \citep{2010Natur.468..669B,2011ApJ...743...92B,2015A&A...576L..11S,2016A&A...596A..47S,2017MNRAS.468.3123S,2016A&A...587A..67L,2016ApJ...832..191N,2018Natur.557..526N,Nikolov2021,2017MNRAS.467.4591G,2020MNRAS.494.5449C,2020MNRAS.497.5155W}.

Observations to date reveal that most exoplanet atmospheres exhibit some level of haze/cloud in their atmosphere \citep{2016Natur.529...59S} and scattering slopes have been detected across the whole continuum, from low \citep[HATS-8b,][]{2020AJ....159....7M} to intermediate  \citep[WASP-43b,][]{2020AJ....159...13W} surface gravities, and from warm Saturns \citep[HAT-P-18b,][]{2017MNRAS.468.3907K} to ultra-hot Jupiters \citep[WASP-12b,][]{2016Natur.529...59S}. While in some cases the slopes can be a result of stellar activity \citep[e.g.][]{2014ApJ...791...55M,2017ApJ...834..151R}, most can be explained with physical properties and composition of the planetary atmosphere. H$_2$ scattering, metal clouds and products from photochemical reactions \citep[e.g.][]{2008A&A...481L..83L,2017MNRAS.471.4355P,2020ApJ...895L..47O} provide a viable explanation of some of the observed negative slopes, with each new observation helping elucidate the mechanisms shaping these atmospheres.

In this paper, we report the optical transmission spectrum of the low density hot-Jupiter WASP-88b, obtained with the VLT FORS2 instrument. Our observations are part of a large VLT exoplanet survey that aims to explore the diversity of exoplanetary atmospheres by contributing to the growing catalogue of planetary atmospheres observed in transmission. FORS2 observations will also provide highly complementary optical transmission spectra for the upcoming James Webb Space Telescope \citep[JWST,][]{2006SSRv..123..485G}. In addition, we report updated physical properties of the system from observations with the Transiting Exoplanet Survey Satellite  \citep[TESS,][]{2015JATIS...1a4003R}.

This paper is organized as follows: Section~\ref{sec:observations_and_data_reduction} details the observations and reductions. Section~\ref{sec:analysis} presents the white and spectroscopic light curve analyses. Section~\ref{sec:transmission_spectrum} summarizes the transmission spectrum and Sections~\ref{sec:discussion} and~\ref{sec:conclusion} present our results, discussion and conclusions.

\subsection{The WASP-88b system}
\label{sec:the_system}

WASP-88b is a transiting hot Jupiter \citep{2014A&A...563A.143D} with a mass of 0.52~M$_\mathrm{Jup}$ and a radius of 1.63~R$_\mathrm{Jup}$, which corresponds to a low surface gravity of 4.8~m~s$^{-2}$ and mean density of 0.11~$\rho_{\mathrm{Jup}}$ (see Section\,\ref{sec:analysis:tess}). The planet moves on a 4.95~d circular orbit around an F6 dwarf of mass 1.29~M$_{\sun}$ and radius 1.93~R$_{\sun}$. The star has an effective temperature of 6450~K and a solar metallicity \citep[SWEET-Cat\footnote{\texttt{www.astro.up.pt/resources/sweet-cat/}},][]{2017A&A...600A..69A}, resulting in an equilibrium temperature for the planet of 1740~K. Simulated cloud-free atmospheres around this temperature predict  a transition from pressure-broadened Na and K lines to TiO and VO in the optical transmission spectrum \citep{2008ApJ...678.1419F,2010ApJ...709.1396F}. The low planet surface gravity suggests an extended atmosphere with a pressure scale height of 1300 km, assuming a mean molecular weight of 2.3 amu. This makes WASP-88b a good target for atmospheric characterisation via transmission spectroscopy. The expected atmospheric signal, i.e. the variation in the transit depth over one pressure scale height, is estimated to be $\Delta\delta=2H R_\mathrm{p}/R^2_*\approx170$ ppm \citep{2010arXiv1001.2010W}.

Magnetic activity in a transiting planet host star can affect the transit shape and transmission spectrum due to the presence of dark spots on the stellar surface. In the case of WASP-88 this can be safely ignored because there is evidence that the star is inactive. At 6450~K the star is too hot to show significant spot activity, and there is no rotational modulation in its long-term light curve to a limit of 1\,mmag \citep{2014A&A...563A.143D}. We performed a period analysis on the TESS data (see below) with the transits removed, which showed no signals to a limit of 0.3\,mmag. We also have spectra of the calcium H and K lines which show a deep line core and no trace of chromospheric emission (data currently under analysis).

The WASP-88 system is also known to have a faint nearby star. High-contrast imaging with the SPHERE instrument on the VLT showed a companion at an angular distance of 3.350 $\pm$ 0.015$\arcsec$ that is fainter by 7.60 $\pm$ 0.53 mag in the $K$-band than the planet host star \citep{2020A&A...635A..73B}. The contamination from nearby companions can affect the shape of a measured transmission spectrum, e.g.\ the emblematic case of WASP-103 \citep{2016MNRAS.463...37S,2017A&A...606A..18L,2020MNRAS.497.5155W}. However, for WASP-88, the faintness of the companion star means its effect on our observations is negligible.

\section{Data acquisition and reduction}
\label{sec:observations_and_data_reduction}

\begin{figure*}
	\centering
	\includegraphics[width=\textwidth,height=\textheight,keepaspectratio]{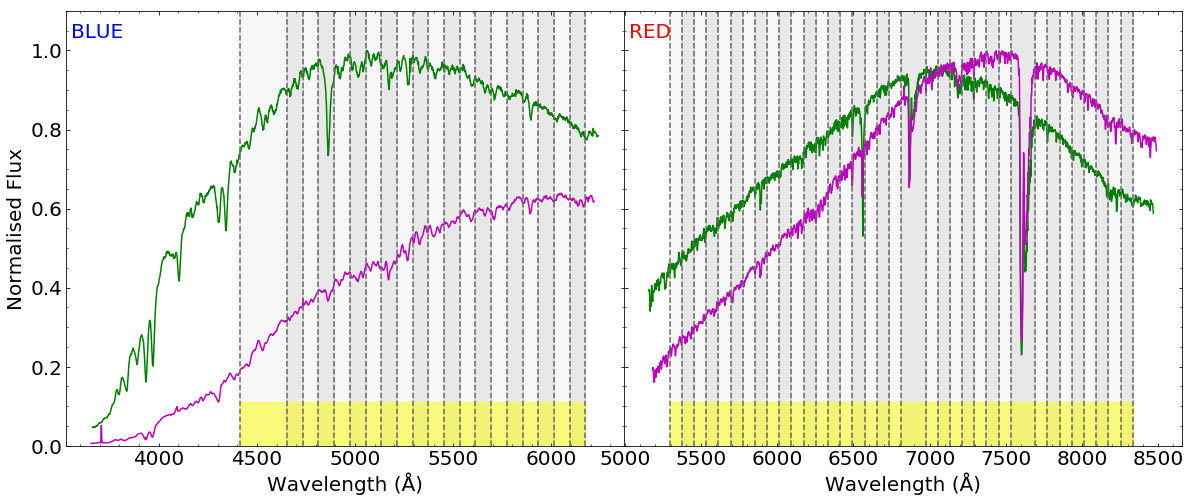}
	\caption{Normalised example spectra from the FORS2 600B (left) and 600RI (right) grism transits. The green line indicates WASP-88 while magenta represents the reference star. The grey bands define the range of each wavelength bin used in our analysis. The yellow regions indicate the spectral range for the white light curves.}
	\label{fig:spectra}
\end{figure*}

\subsection{VLT FORS2}

We observed two complete transits of WASP-88b using the VLT low resolution spectrograph FORS2 \citep{1998Msngr..94....1A} mounted on the Unit Telescope 1 (UT1) Cassegrain focus at the European Southern Observatory (ESO) in  Paranal, Chile. These time series observations were performed during the nights of 2017 August 19th and 24th as part of program 199.C-0467 (PI: Nikolov). We observed only one suitably bright reference star in the FORS2 field of view and took advantage of the multi-object spectroscopy mode to simultaneously acquire spectra of our target and the reference star, which is known as 2MASS J20381555$-$4829215 and is located at an angular separation of 2.7$\arcmin$. The two spectra were collected separately by the two CCDs of the red detector. We made use of the MXU mode and devised a custom built mask with slits of size 22$\arcsec$x132$\arcsec$ for both stars. The broad slits were applied to minimize differential slit light losses due to seeing fluctuations. We also optimised the duty cycle to a readout time of $\sim30\sec$ by selecting a binning of 2x2. In addition, due to the long duration of the observations ($>$8 hours), we used a range of integration times to adjust for variations in flux. This is not a problem for the detectors, as FORS2 CCDs are linear up to saturation.

We employed the dispersive element GRIS600B (hereafter blue or 600B) to record the first transit, which covers the wavelength range from 3300 to 6200~\AA. The sky was clear throughout the observation with low level atmospheric turbulence keeping seeing above 1$\arcsec$ (0.82$\arcsec$-1.8$\arcsec$) during most of the campaign. We followed the target as it ascended from an airmass of 1.38 to an airmass of 1.09 and then descended to an airmass of 2.21. In total, 206 spectra were collected within a period of 8 hours and 12 $\min$ and with integration times of 30, 100 or 120 $\sec$.

The second transit was observed using the dispersive element GRIS600RI (hereafter red or 600RI) in conjunction with blocking filter GG435, which isolates the first spectral order. The red grating was utilised to cover the spectral region between 5200 and 8400\AA. During this night, the sky was mostly clear and photometric conditions were observed in the early hours of August 25th. The target was tracked as it ascended from an airmass of 1.39 to an airmass of 1.09 and then descended to an airmass of 2.22. The seeing remained below 1$\arcsec$ most of the night, reaching a low of 0.4$\arcsec$ and a high of 1.94$\arcsec$. Both the target and the reference star were monitored for 8 hours and 19 $\min$ resulting in 494 exposures with integration times between 25 and 80 $\sec$.

We then proceeded with the reduction of the data by employing a custom-built, IDL-based pipeline commonly used in FORS2 analyses \citep[e.g.][]{2016ApJ...832..191N,2018Natur.557..526N,2020MNRAS.494.5449C}. Our initial step was to subtract the bias frames and perform flat field corrections to the raw images. However, we found no significant improvements to the data and opted to carry out our analysis without these corrections. To extract the aperture of each spectrum we made use of the APALL subroutine included in the IRAF package and performed simple box summation. We identified an aperture radius of 21 pixels to be the best solution that minimises the scatter in the out-of-transit data for both data sets. We also defined sky background regions on both sides of each spectrum and subtracted the median count values from the spectral trace. These regions were located 30 to 80 pixels away from the spectral peak. This step is important, as it also removes the bias level. Normalised example spectra of WASP-88 and the comparison star are shown in Figure~\ref{fig:spectra}.

To establish a wavelength solution for the stellar spectra, we used an emission lamp at the end of each observation. We applied a mask that is very similar to the one used during stellar tracking but with a slit width of 1\arcsec. From this, we determined a wavelength solution for each spectrum by estimating the centroid location of the most prominent lines through Gaussian fitting and then performing low-order Chebyshev polynomial fits to the computed centroids. We accounted for minor sub-pixel displacements in the dispersion direction during each observation by cross-correlating the extracted spectra against a reference Doppler-corrected rest frame.

\subsection{TESS}
\label{sec:data:tess}

WASP-88 was recently observed using the TESS satellite \citep{2015JATIS...1a4003R} in short cadence, in Sector 27. This light curve contains 16\,511 points, which cover 24.4\,d at a sampling rate of 120.1\,s, with a short break near the middle for the transmission of data back to Earth. Four complete transits occur within these data, and there are no gaps in sampling within them. There is partial coverage of one more transit, which we neglected.

We downloaded the TESS data for WASP-88 from MAST\footnote{\texttt{https://mast.stsci.edu/portal/Mashup/Clients/Mast/\allowbreak Portal.html}} and extracted the PDC fluxes \citep{2016SPIE.9913E..3EJ}, imposing a requirement that the quality flag must equal zero. We then removed all data further than 1.5 transit durations from the midpoint of a transit, leaving a total of 1998 datapoints. We normalised each transit to unit flux by fitting and dividing out a straight line to the data either side of each transit.

\section{Analysis}
\label{sec:analysis}

\subsection{TESS}
\label{sec:analysis:tess}

The previous analysis of WASP-88 \citep{2014A&A...563A.143D} was based on five transit light curves, of which only two covered all four contact points. The existence of the TESS data (Sec.\,\ref{sec:data:tess}) allows this situation to be improved. We modelled the TESS data with the {\sc jktebop} code \citep{2013A+A...557A.119S} following the precepts of the \textit{Homogeneous Studies} project \citep[][and references therein]{2012MNRAS.426.1291S}. After an initial fit to determine the orbital ephemeris, we condensed the data by sorting according to orbital phase and binning each successive five datapoints together. This phase-binning process yielded 400 binned datapoints with an effective sampling rate of 150\,s.

This light curve was modelled using {\sc jktebop}, with the parameters of the fit being the sum of the fractional radii ($r_{\rm *}+r_{\rm p}$ where $r_{\rm *}=\frac{R_{\rm *}}{a}$ and $r_{\rm p}=\frac{R_{\rm p}}{a}$, $R_{\rm *}$ is the radius of the star, $R_{\rm p}$ is the radius of the planet, and $a$ is the semi-major axis of the relative orbit.), the ratio of the radii ($k = \frac{r_{\rm p}}{r_{\rm *}}$), the orbital inclination ($i$), the out-of-transit light level and the phase of mid-transit. A circular orbit was assumed based on the results of \citet{2014A&A...563A.143D}. Although there is a nearby star included in the PSF of the TESS data \citep{2020A&A...635A..73B} we assumed that contaminating light was negligible because \citet{2020A+A...635A..74S} found that stars more than 3\,mag fainter have a negligible effect on analyses such as the current one. For reference, we find that the fractional light contribution of this object in the TESS band is only $(9.1 \pm 4.1)\times 10^{-6}$ using the calculation method from \citet{2020A+A...635A..74S}.

Limb darkening was implemented using the quadratic, square-root, logarithmic and cubic laws \citep{2008MNRAS.386.1644S}. Fits were obtained for two approaches for each law: with both coefficients fixed at theoretical values and with one coefficient fitted whilst the other was held fixed. The values of the limb darkening coefficients were taken from \citet{2017A+A...600A..30C}. 

\begin{figure} \centering
	\includegraphics[width=\columnwidth]{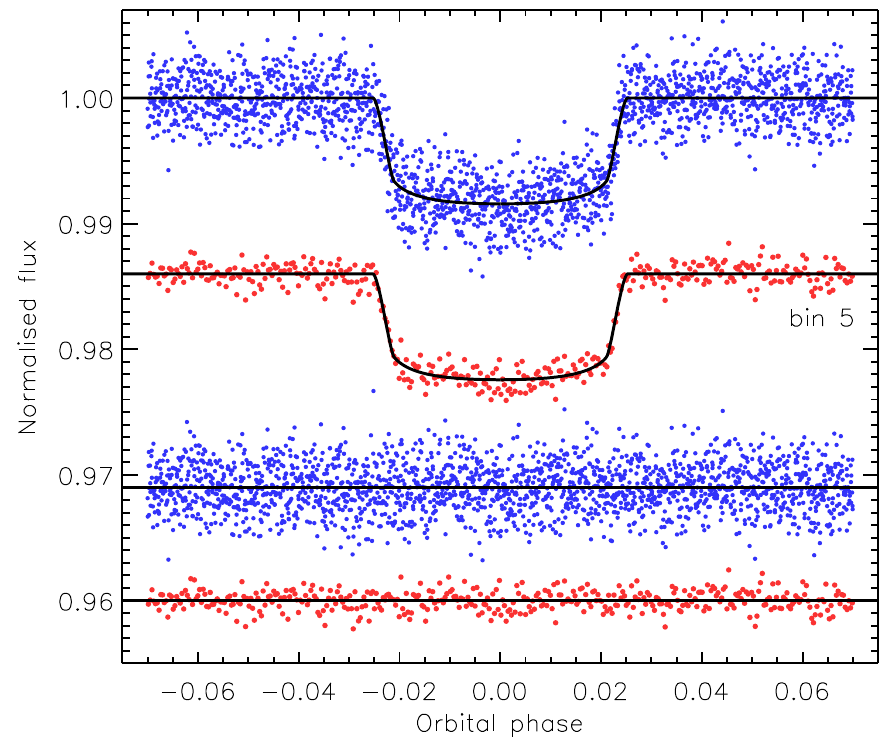} 
	\caption{TESS light curve (blue circles) and phase-binned light curve (red circles) of WASP-88 compared to the best fits found using the {\sc jktebop} code (black solid lines). The residuals of the fits are shown offset to the base of the figure. All data are shown versus orbital phase for clarity. The phase-binned data are those used to determine the physical properties of the system, whereas the unbinned data were utilised for determination of the orbital ephemeris.} 
	\label{fig:tess}
\end{figure}

Uncertainties in the fitted parameters were calculated using Monte Carlo and residual-permutation algorithms \citep{2008MNRAS.386.1644S}, which have been found to be consistent with errorbars returned from several types of MCMC analyses \citep{2020MNRAS.498..332M}. An additional contribution to the uncertainties is the variation between fits with different limb darkening laws, and this was assessed and added in quadrature to the larger of the Monte Carlo and residual-permutation errorbars. The final photometric parameters are given in Table\,\ref{tab:tess}. A plot of the TESS data and best fit is given in Fig.\,\ref{fig:tess}.

The physical properties of the system were determined using the $r_{\rm *}$, $r_{\rm p}$, $i$ and orbital period from the \textsc{jktebop} analysis, the effective temperature and metal abundance from \citet{2017A&A...600A..69A}, and the velocity amplitude of the star ($K_{\rm *} = 53.4^{+6.8}_{-6.6}$\,m\,s$^{-1}$) from \citet{2014A&A...563A.143D}. Using these quantities, we identified the value of the velocity amplitude of the planet ($K_{\rm p}$) that gave the best agreement between the measured $T_{\rm eff}$ and radius of the star compared to the predictions of five different sets of theoretical stellar evolutionary models \citep[see][]{2010MNRAS.408.1689S}. Uncertainties in all the input parameters were propagated by a perturbation analysis and added in quadrature for each output parameter.

The final physical properties of the WASP-88 system are given in Table\,\ref{tab:tess}. Systematic uncertainties were measured from the scatter of each output parameter over the results for the five different sets of theoretical stellar models used. The final parameters are in good agreement with those found by \citet{2014A&A...563A.143D}, although we find a modestly smaller radius of the star and thus planet. The measured $T_{\rm eff}$ of the star is difficult to match to its density obtained from the light curve, leading to a significant systematic uncertainty in the stellar mass. We quote the modified equilibrium temperature as defined by \citet{2010MNRAS.408.1689S}.

\begin{table} \centering
\caption{Parameters of WASP-88 obtained from analysis of the TESS light curve. When two sets of errorbars are given they refer to the random and systematic uncertainties, respectively. The $T_{\rm eff}$ and [Fe/H] come from \citet{2017A&A...600A..69A}.} \label{tab:tess}
\begin{tabular}{lc}
\hline
\hline
Parameter & Value \\
\hline
\multicolumn{2}{l}{\textit{Light curve parameters}} \\
Orbital period (d) & 4.9540045 $\pm$ 0.0000020 \\
Time of transit (BJD/TDB) & 2456474.73154 $\pm$ 0.00087 \\
$r_{\rm *}+r_{\rm p}$ & $0.1577\,^{+0.0099}_{-0.0025}$ \\[2pt]
$k$ & $0.0869\,^{+0.0010}_{-0.0012}$ \\[2pt]
$i$ (degrees) & $89.7\,^{+0.3}_{-2.6}$ \\[2pt]
$r_{\rm *}$ & $0.1451\,^{+0.0088}_{-0.0024}$ \\[2pt]
$r_{\rm p}$ & $0.01261\,^{+0.00091}_{-0.00024}$ \\
\hline
\multicolumn{2}{l}{\textit{Physical properties}} \\
$T_{\rm eff}$ (K) & 6450 $\pm$ 61 \\
\mbox{[Fe/H]} & 0.03 $\pm$ 0.04 \\
Stellar mass (M$_\odot$) & $1.288\,^{+0.029}_{-0.020}\,^{+0.020}_{-0.025}$ \\[2pt]
Stellar radius (R$_\odot$) & $1.93\,^{+0.12}_{-0.03}\,^{+0.01}_{-0.01}$ \\[2pt]
Stellar log $g$ (c.g.s.) & $3.976\,^{+0.015}_{-0.051}\,^{+0.002}_{-0.003}$ \\[2pt]
Stellar density ($\rho_\odot$) & $0.179\,^{+0.009}_{-0.029}$ \\[2pt]
System age (Gyr) & $1.7\,^{+0.2}_{-0.4}\,^{+0.2}_{-0.4}$ \\[2pt]
Planet mass (M$_{\rm Jup}$) & $0.520\,^{+0.067}_{-0.066}\,^{+0.005}_{-0.007}$ \\[2pt]
Planet radius (R$_{\rm Jup}$) & $1.63\,^{+0.12}_{-0.03}\,^{+0.01}_{-0.01}$ \\[2pt]
Planet surface gravity (m~s$^{-2}$) & $4.84\,^{+0.64}_{-0.88}$ \\[2pt]
Planet mean density ($\rho_{\rm Jup}$) & $0.112\,^{+0.016}_{-0.026}$ \\[2pt]
Planet equilibrium temperature (K) & $1737\,^{+54}_{-15}$ \\[2pt]
Semimajor axis (au) & $0.06189\,^{+0.00048}_{-0.00032}\,^{+0.00031}_{-0.00040}$ \\
\hline
\end{tabular}
\end{table}

\subsection{VLT FORS2}

We produced two white light curves (one from each dataset) and 56 spectroscopic light curves (20 from the first night and 36 from the second). The two observations covered the wavelength range from 4413 to 6173 \AA\ and from 5293 to 8333 \AA. The spectroscopic channels were sorted into narrow bins of 80 and, in some cases, 160 or 240 \AA\ following most of the same spectral bands presented in \citet{2018Natur.557..526N}. The wavelength region between 4013 and 4413 \AA\ was excluded from the rest of our analysis due to low signal to noise ratio. To correct for atmospheric effects, including variations in extinction due to shifting airmass and contamination from telluric lines, we divided the flux of the target by the flux of the reference star.

\subsubsection{White light curves}
\label{sec:white_analysis}

To model the white light curves, we treated the data as a Gaussian process (GP) by utilising the Python GP package \texttt{george} \citep{2015ascl.soft11015F}. Under the GP condition, the data are described by a multivariate normal probability distribution $p$, which consists of a mean function $T$ that defines the deterministic input from the transit and a covariance matrix $K$ that describes the stochastic noise component \citep{2012MNRAS.419.2683G,2014MNRAS.445.3401G}: 
\begin{equation}p(\mathcal{D} \mid \theta, \phi)=\mathcal{N}(f \mid T(t, \theta), K)\end{equation}

Here, $\mathcal{D}$ represents the data, $\mathcal{N}$ signifies the multivariate normal distribution, $f$ is a vector of the relative flux measurements, $\theta$ are the transit model parameters and $\phi$ is the hyperparameter vector of the kernel function. Vector $t$ specifies the central exposure times after their conversion from Modified Julian Dates (MJD) to Barycentric Julian Dates (BJD) using the Python library \texttt{barycorrpy} \citep{2018RNAAS...2a...4K}. This converter offers a clock correction, a geometric correction and an Einstein correction \citep[for a detailed explanation, see][]{2010PASP..122..935E}.

For the mean function, we employed the open source package \texttt{batman} \citep{2015PASP..127.1161K} to compute the model transit light curves. This code is able to compute model transits from a wide range of stellar limb darkening laws. During the parametrisation procedure, we adopted the quadratic limb darkening law \citep{1950HarCi.454....1K} as it is computationally less demanding than more complicated laws and has been proven to deliver relatively accurate results in numerous studies of planetary atmospheres \citep[e.g.][]{2016ApJ...832..191N, 2017MNRAS.467.4591G}. In this case, \texttt{batman} follows the analytic algorithms described in \citep{2002ApJ...580L.171M}. To validate the reliability of our choice, we performed the same analysis using the more complex four-parameter non-linear law \citep{2000A&A...363.1081C}. The two approaches were found to produce consistent results and agree at the 1$\sigma$ level. 

To model light curve systematics, we chose the Mat\'{e}rn 3/2 kernel \citep[for a more in-depth explanation see][]{2012RSPTA.37110550R}. Our choice is motivated by the significant light curve systematics and the fact that this kernel can be differentiated a finite number of times making it less smooth than the squared exponential kernel (infinitely differentiable), which is the other common choice in such analyses. Furthermore, \citet{2013MNRAS.428.3680G} found through empirical methods that the Mat\'{e}rn 3/2 kernel performs better than other kernels in light curves with time-dependent noise such as the ones from the blue and red data sets (the rate of change of the rotator angle is a function of time). The covariance matrix is then defined as: 
\begin{equation}K =\xi^{2}\left(1+\sqrt{3} D_{n m}\right) \exp (-\sqrt{3} D_{n m})+\delta_{n m} \left(\sigma_{n}\sigma_{\alpha}\right)^{2},\end{equation}
where $\xi$ is the height scale or correlation amplitude, $\delta_{n m}$ is the Kronecker delta, $\sigma_{n}$ are the spectrophotometric uncertainties, determined from an assumption of pure photon noise, and $\mathrm{D}_{nm}$ is given by:
\begin{equation}D_{nm}=\sqrt{\sum_{\nu=1}^{N}\left(\frac{\left(\hat{w}_{\nu,n}-\hat{w}_{\nu,m}\right)^{2}}{\tau_{w_\nu}^{2}}\right)},\end{equation}
where $\tau_{w_\nu}$ are the length scale parameters for each external systematic variable $\hat{w}_{\nu}$ used. The hats here indicate that the variables are standardised (i.e. their values are set on the same scale by subtracting the mean and dividing by the standard deviation). We also fit for a multiplicative factor $\sigma_{\alpha}$, which rescales our photon noise uncertainties to more realistic values and is the same for all flux measurements.

We considered a variety of auxiliary systematic variables for the kernel functions of our datasets, including airmass, positional drifts, rotator angle changes, FWHM, sky background, and ambient pressure and temperature, and determined that each transit light curve is best described by a different set of systematics (see Figure~\ref{fig:auxiliary_variables} for the real time trends in selected ancillary variables). More specifically, after an inspection of the posterior distributions and the fitted light curves, we concluded that the influence of physical parameters in the observed blue light curve is negligible. We therefore chose to assume only time $\hat{w}_{\nu}=\hat{t}$ as a detrending factor in this case. However, for the red data, we found that a combination of shifts in the dispersion ($x$) and cross-dispersion ($y$) directions and the rate of change of the rotator angle ($z$) significantly improve the shape of the posterior distribution and increase the precision of our estimated parameters whereas time alone shows a strong correlation with the transit depth. In addition, the rotator angle could still be affected by inhomogeneities in the spatial transmission of the longitudinal atmospheric dispersion corrector (LADC) despite a recent fix \citep{2016SPIE.9908E..2BB}. Thus, we opted to use three systematic variables: $\hat{w}_{\nu}=(\hat{w}_1, \hat{w}_2, \hat{w}_3)=(\hat{x}, \hat{y}, \hat{z})$. We also examined the impact of a linear trend described by a function of time or airmass. We observed no significant divergence from the simpler mean function models and so a polynomial of this kind was excluded from the rest of our analysis.

We allowed four transit parameters $\theta=(t_0, R_\mathrm{p}/R_*, a/R_*, i)$, a white noise term $\sigma_{\alpha}$ and a set of hyperparameters $\phi=(\xi, \tau_{w_\nu})$ to vary freely in our fit for each white light curve. The planet-to-star radius ratio $R_\mathrm{p}/R_*$, the semi-major axis to stellar radius ratio $a/R_*$ and the orbital inclination $i$ were initially placed to the values from Table~\ref{tab:tess} with the time of mid-transit set to the expected value according to the ephemeris given in the same table. We computed the theoretical values for the two limb darkening coefficients $u_1$ and $u_2$ using the Stagger-grid \citep{2015A&A...573A..90M}. The 3D model stellar atmosphere, from which the coefficients were derived, was generated by taking into account the closest values to the metallicity, surface gravity and effective temperature reported in \citet{2014A&A...563A.143D}. We considered cases where one or both limb darkening coefficients vary freely in the fit but noticed this leads to poor constraints on their values. Consequently, we opted to fix the two coefficients to their theoretical values. We also fixed the eccentricity to 0 (assuming a circular orbit) and the period to the value obtained from TESS. Finally, we applied log-uniform priors to the hyperparameters and uniform priors to all other parameters.

The optimised transit and kernel parameters and their uncertainties were retrieved from a Markov-Chain Monte Carlo sampling process. We employed the affine invariant ensemble sampler \citep{2010CAMCS...5...65G} from the python implementation \texttt{emcee} \citep{2013PASP..125..306F} to marginalise the posterior distribution. This method was chosen over standard MCMC algorithms as it explores the parameter space quickly and efficiently from a set of walkers that steadily progress to higher likelihoods in the probability distribution through random linear combinations with other members of the ensemble. We opted to adopt a group of 150 walkers and determined the best-fit result in two three-stage iterations. In stage 1, we initialised our walkers to be close to the literature estimates (for the transit parameters) or some arbitrary values close to ones from other targets (for the hyperparameters) and ran an MCMC chain of 500 steps. We then re-initialised our walkers to a tight region around the position of the walker with the best likelihood and performed a new run with the same amount of steps. This stage was included to accelerate convergence towards the optimal solution. The final production chain was executed in 5000 steps and the median results from the marginalised posterior distribution of the second iteration are shown in Table~\ref{tab:wasp88_white}. The full distributions can be seen in Figures~\ref{fig:blue_corner} and~\ref{fig:red_corner}.

\begin{figure*}
	\centering
	\includegraphics[width=\textwidth,height=\textheight,keepaspectratio]{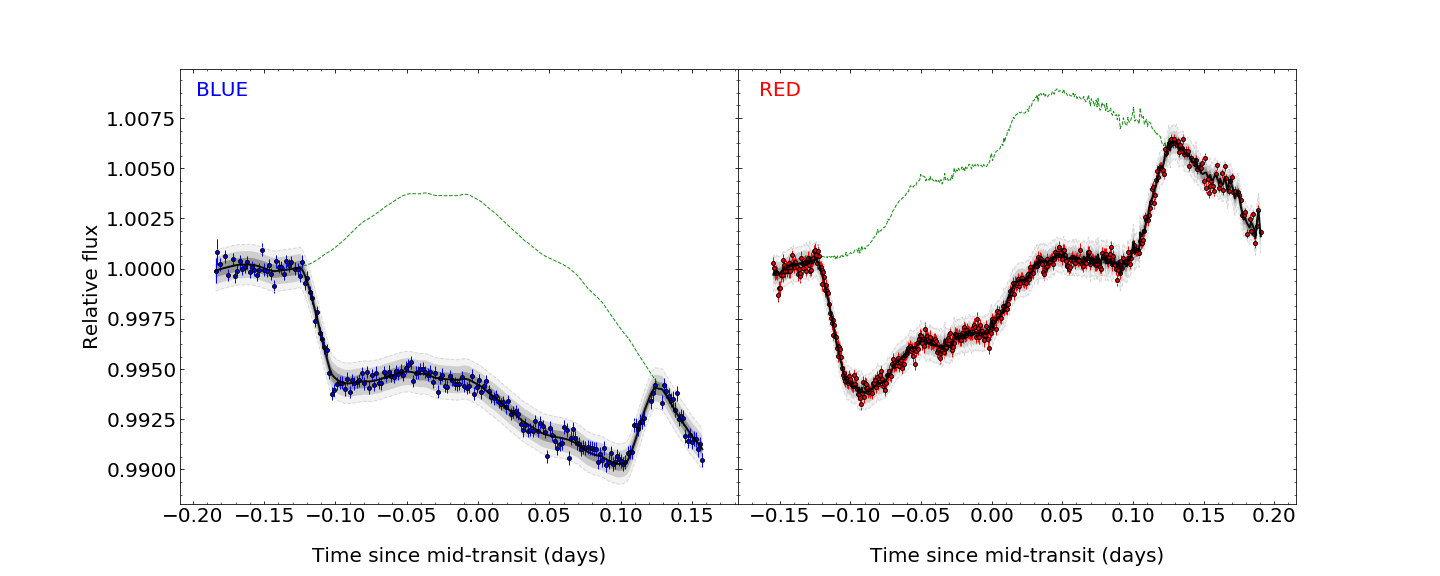}
	\caption{VLT FORS2 white-light transit light curves of WASP-88b. The black line indicates the GP model and the green dashed line displays the systematics model. The transparent grey regions show the 1$\sigma$, 2$\sigma$ and 3$\sigma$ error of the residuals (from darker to lighter shades), with error bars indicating the spectrophotometric uncertainties.}
	\label{fig:both_white}
\end{figure*}

We performed a second fit to both light curves by fixing $t_0$ to the values recovered from the first fit, and $a/R_*$ and $i$ to their calculated weighted means. Any data points that deviated from the first GP fit by more than three times the standard deviation of the residuals were discarded in our analysis. The new model fit for the blue and red light curves, along with the noise component, is shown in Figure~\ref{fig:both_white}.

\begin{table}
\centering
\caption{Parameters from the white transit light curve analysis. $^{*}$For the time of mid-transit, we use the expected values from the ephemeris given in Table~\ref{tab:tess} and we assume these to be 0 for the prior ranges.}
\label{tab:wasp88_white}
\begin{tabular}{lcc}
\hline
\hline
Parameter & Value & Prior\\
\hline
$P$ (d) & 4.9540045 (fixed) & \\
$e$ & 0 (fixed) & \\
 & & \\
blue (1st fit) & & \\
$t_0$ (BJD/TDB) & $2457985.70345^{+0.00070}_{-0.00069}$ & $\mathcal{U}$(-0.01,0.01)$^{*}$\\[2pt]
$R_\mathrm{p}/R_*$ & $0.0885^{+0.0046}_{-0.0044}$ & $\mathcal{U}$(0.03,0.15)\\[2pt]
$a/R_*$ & $6.22^{+0.24}_{-0.22}$ & $\mathcal{U}$(4,9)\\[2pt]
$i$ (degrees) & $86.07^{+0.87}_{-0.73}$ & $\mathcal{U}$(80,90)\\[2pt]
$u_1$ & 0.327 (fixed) & \\
$u_2$ & 0.363 (fixed) & \\
$\ln\alpha$ & $-8.5^{+2.0}_{-1.4}$ & $\mathcal{U}$(-20,15)\\[2pt]
$\ln\tau_{\mathrm{t}}$ & $2.9^{+1.5}_{-1.1}$ & $\mathcal{U}$(-15,15)\\[2pt]
$\sigma_{\mathrm{\alpha}}$ (from 1st iteration) & $4.31^{+0.23}_{-0.21}$ & $\mathcal{U}$(0,10)\\[2pt]
 & & \\
red (1st fit) & & \\
$t_0$ (BJD/TDB) & $2457990.65735 \pm 0.00045$ & $\mathcal{U}$(-0.01,0.01)$^{*}$\\[2pt]
$R_\mathrm{p}/R_*$ & $0.0868^{+0.0032}_{-0.0035}$ & $\mathcal{U}$(0.03,0.15)\\[2pt]
$a/R_*$ & $6.62^{+0.16}_{-0.17}$ & $\mathcal{U}$(4,9)\\[2pt]
$i$ (degrees) & $87.68^{+0.98}_{-0.74}$ & $\mathcal{U}$(80,90)\\[2pt]
$u_1$ & 0.193 (fixed) & \\
$u_2$ & 0.364 (fixed) & \\
$\ln\alpha$ & $-10.33^{+1.12}_{-0.78}$ & $\mathcal{U}$(-20,15)\\[2pt]
$\ln\tau_{\mathrm{x}}$ & $6.2^{+1.3}_{-1.1}$ & $\mathcal{U}$(-15,15)\\[2pt]
$\ln\tau_{\mathrm{y}}$ & $1.14^{+0.81}_{-0.60}$ & $\mathcal{U}$(-15,15)\\[2pt]
$\ln\tau_{\mathrm{z}}$ & $2.64^{+0.95}_{-0.77}$ & $\mathcal{U}$(-15,15)\\[2pt]
$\sigma_{\mathrm{\alpha}}$ (from 1st iteration) & $3.52 \pm 0.13$ & $\mathcal{U}$(0,10)\\[2pt]
 & & \\
Weighted mean: & & \\
$a/R_*$ & $6.49 \pm 0.14$ & \\[2pt]
$i$ (degrees) & $87.07^{+0.59}_{-0.50}$ & \\[2pt]
 & & \\
blue (2nd fit) & & \\
$R_\mathrm{p}/R_*$ & $0.0882^{+0.0037}_{-0.0036}$ & $\mathcal{U}$(0.03,0.15)\\[2pt]
$\ln\alpha$ & $-8.5^{+2.1}_{-1.4}$ & $\mathcal{U}$(-20,15)\\[2pt]
$\ln\tau_{\mathrm{t}}$ & $2.9^{+1.5}_{-1.1}$ & $\mathcal{U}$(-15,15)\\[2pt]
$\sigma_{\mathrm{\alpha}}$ (from 1st iteration) & $0.83^{+0.05}_{-0.04}$ & $\mathcal{U}$(0,10)\\[2pt]
 & & \\
red (2nd fit) & & \\
$R_\mathrm{p}/R_*$ & $0.0858^{+0.0031}_{-0.0033}$ & $\mathcal{U}$(0.03,0.15)\\[2pt]
$\ln\alpha$ & $-10.56^{+0.96}_{-0.68}$ & $\mathcal{U}$(-20,15)\\[2pt]
$\ln\tau_{\mathrm{x}}$ & $5.9^{+1.2}_{-1.0}$ & $\mathcal{U}$(-15,15)\\[2pt]
$\ln\tau_{\mathrm{y}}$ & $0.81^{+0.70}_{-0.55}$ & $\mathcal{U}$(-15,15)\\[2pt]
$\ln\tau_{\mathrm{z}}$ & $2.17^{+0.88}_{-0.73}$ & $\mathcal{U}$(-15,15)\\[2pt]
$\sigma_{\mathrm{\alpha}}$ (from 1st iteration) & $0.88 \pm 0.03$ & $\mathcal{U}$(0,10)\\[2pt]
\hline
\end{tabular}
\end{table}

\subsubsection{Spectroscopic light curves}
\label{sec:spectroscopic_analysis}

A widespread practice in transmission spectroscopy is the application of a common mode correction to the spectrophotometric light curves due to the presence of a wavelength-independent noise component in all the time series of the same data set \citep[e.g.][]{2012MNRAS.426.1663S,2016A&A...587A..67L,2016ApJ...832..191N,2018Natur.557..526N}. We applied the same technique to the spectroscopic light curves obtained from the blue and red data sets. To acquire the common mode factors, we divided the raw white light curves by the median GP model. We then corrected the binned transit light curves by dividing the raw, spectroscopic relative fluxes by this common trend.

The transit modelling of each spectroscopic light curve was performed following the same procedure described in the white light curve analysis. A simple GP kernel of time was employed to express the impact of additional wavelength-dependent systematics. More complex kernels were rejected under the assumption that most of the contribution from physical factors is modelled out during common mode correction. Furthermore, any remaining residuals were found to be fitted well by time alone. Again, the best-fit parameters were determined through a three-stage MCMC likelihood maximisation procedure. The only difference here was that the step size in the last run was reduced to 1000 steps. We found that higher step sizes made no difference to the computed results.

The fixed and retrieved parameters from the second white light curve fits were used as the transit input for the spectroscopic light curves and their values were held fixed throughout the rest of the investigation. An obvious exception is the transmission spectrum parameter $R_\mathrm{p}/R_*$, which was considered to be a free parameter in the modelling of each light curve. The theoretical limb darkening coefficients, in each case, were determined in the same way as in the white light analysis with the quadratic coefficient being kept fixed and the linear one allowed to vary in each fit. We tested configurations where both limb darkening parameters were either fixed or variable and found no improvement to the fit. We also checked a version without common mode correction and observed that the limb darkening coefficients behaved in a similar fashion to the white light by settling at significantly lower values than expected.

\begin{figure*}
\centering
\includegraphics[width=\textwidth,height=0.9\textheight,keepaspectratio]{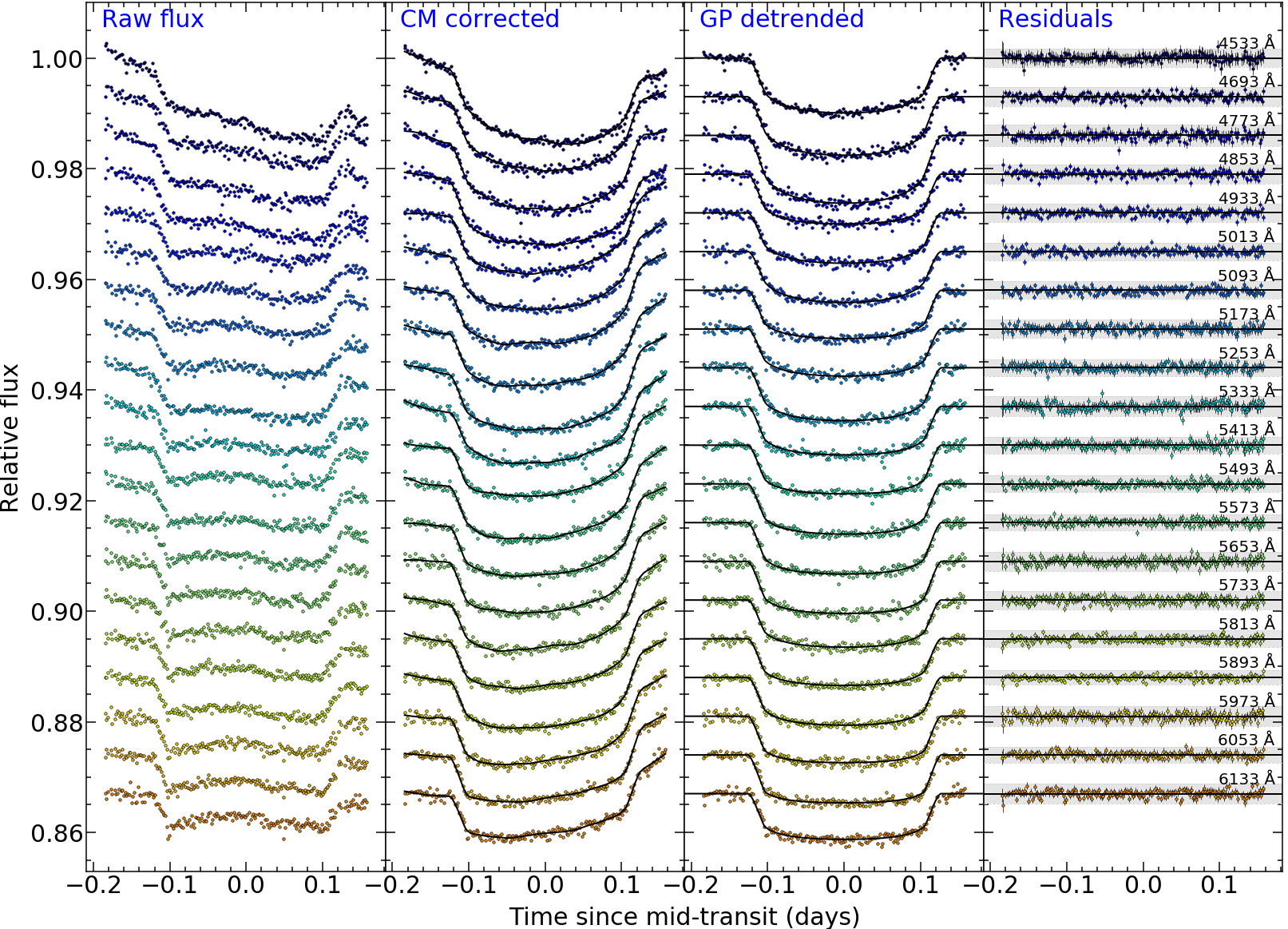}
\caption{Spectroscopic light curves for the blue data set of WASP-88b with lower wavelengths indicated by blue data points and higher wavelengths indicated by brown data points. The light curves are offset from unit flux for clarity. First panel: Raw light curves. Second panel: Common-mode corrected light curves and their respective GP fit. Third panel: Detrended light curves and their respective best-fit transit model. Fourth panel: Residuals from the best-fit model and their 1$\sigma$ spectrophotometric uncertainties (vertical error bars). The transparent grey boxes indicate the 3$\sigma$ region of the residuals.}
\label{fig:blue_spectroscopic}
\end{figure*}

\begin{figure*}
\centering
\includegraphics[width=\textwidth,height=0.9\textheight,keepaspectratio]{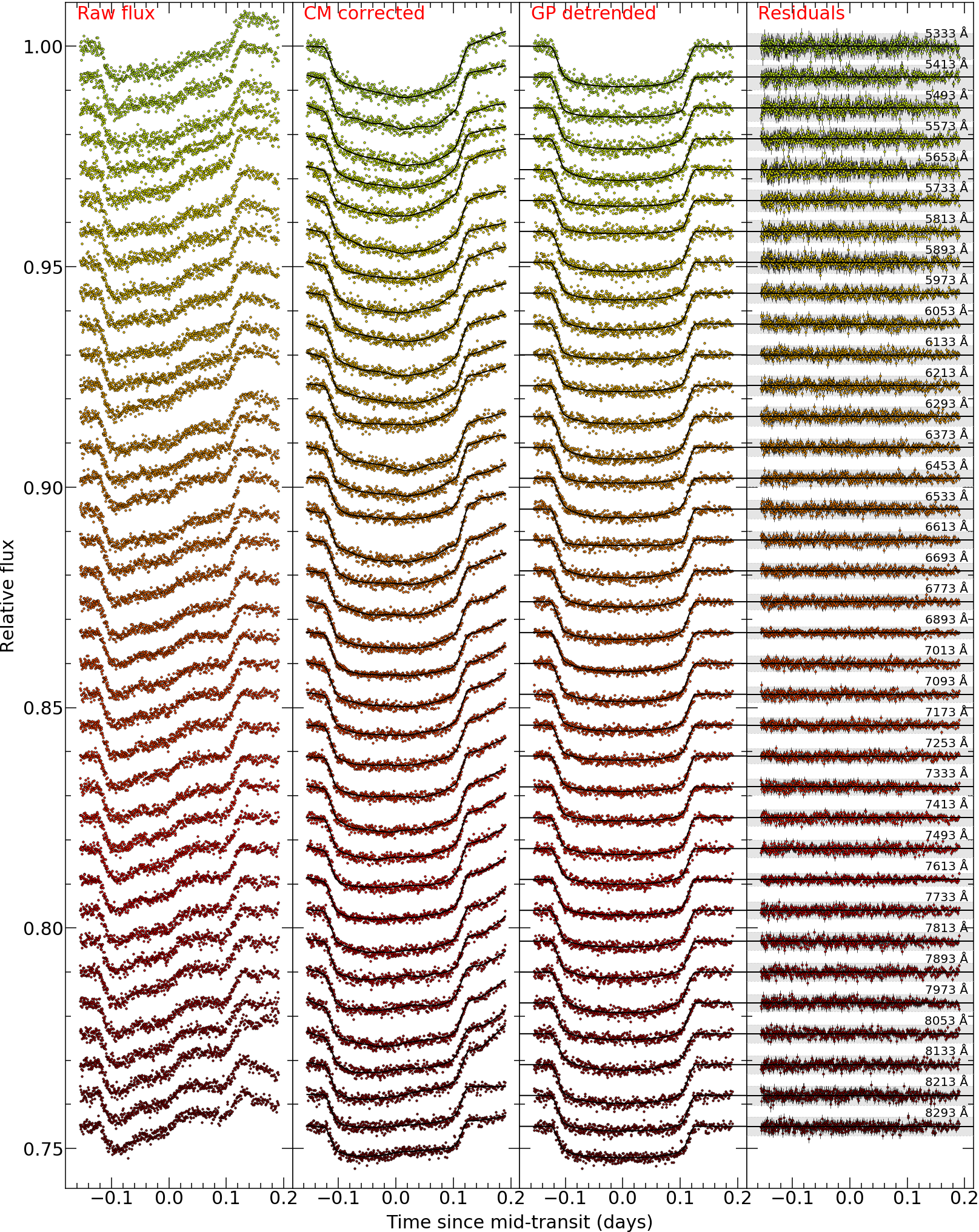}
\caption{Same as Figure~\ref{fig:blue_spectroscopic}, but for the red data set. Lower wavelengths are indicated by light green data points and higher wavelengths by dark red data points.}
\label{fig:red_spectroscopic}
\end{figure*}

Again, a second iteration was performed to reduce the impact of outliers. During this iteration, the free parameters remained the same and the deviating data at the 3$\sigma$ level were removed following an identical approach to the combined light curves. Figures~\ref{fig:blue_spectroscopic} and~\ref{fig:red_spectroscopic} show the various fitting stages of the blue and red spectroscopic light curves and the residuals from the best-fit models. The retrieved values of $R_\mathrm{p}/R_*$ and $u_1$ are reported in Table~\ref{tab:wasp88_spectroscopic}.

\section{Transmission spectrum}
\label{sec:transmission_spectrum}

We constructed the transmission spectrum for the blue and red data sets separately using the planet-to-star radii ratios obtained during the spectroscopic analysis. We found mean values of $R_p/R_* = 0.0884 \pm 0.0035$ (blue) and $R_\mathrm{p}/R_* = 0.0865 \pm 0.0024$ (red), which are in excellent agreement with the $R_\mathrm{p}/R_*$ estimate from TESS (see Table~\ref{tab:tess}). We also identified an offset of magnitude $\Delta (R_\mathrm{p}/R_*) = 0.0006 \pm 0.0026$ in the overlapping region between the two data sets. The offset is small, showcasing the consistency of using the same stochastic technique in both the white and spectroscopic analyses. The somewhat large uncertainty here stems from the residual and system parameter errors and is a natural outcome of the flexible GP fit which also considers unaccounted for noise from instrumental, atmospheric or astrophysical sources, including any residuals from the fixed limb darkening coefficients in the white light curve analysis.

\begin{figure}
\centering
\includegraphics[width=0.5\textwidth,height=0.5\textheight,keepaspectratio]{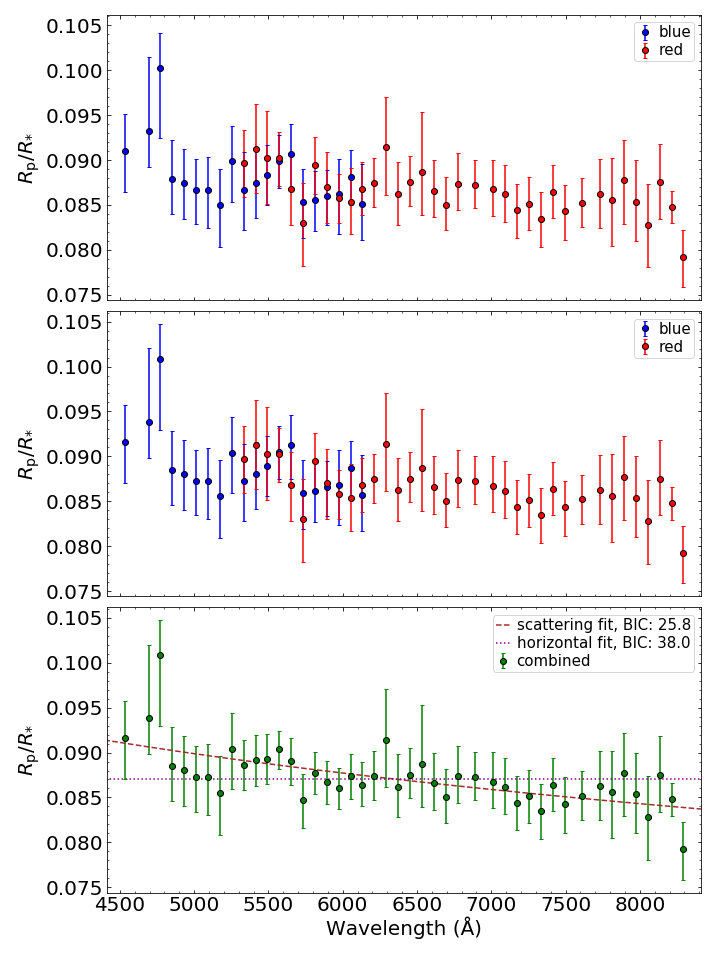}
\caption{The transmission spectrum of WASP-88b before offset correction (top panel), after offset application (middle panel) and after estimation of the weighted mean values in the overlapping region (bottom panel). Blue and red indicate the two different data sets while green indicates the combined final data set. The brown dashed line depicts a fit to the Rayleigh slope whereas the dotted magenta line depicts a straight, horizontal fit.}
\label{fig:transmission_spec}
\end{figure}

Since we do not have any prior spectrophotometric information from previous observations we applied the offset to the blue data set. Our choice to use the red data set as a reference can be justified by the fact that the blue data exhibit more severe atmospheric extinction. In addition, during the white light analysis, we model noise as a function of time only in the blue curves whereas in the red curves we use an assortment of physical parameters\footnote{A preference for more complex GP models in the red data set was also observed in the analysis of WASP-39b \citep{2016ApJ...832..191N}}. Time as a noise factor is quite flexible in its behaviour and can smoothly follow the features within the data. This can have a detrimental effect in the estimation of system parameters, and especially the transit depth, as it can move the values away from truth with the insertion of considerably large error bars. The physical parameters, however, appear more robust to this effect as they may account for some of the spikes observed in the data. After the vertical displacement was implemented, we computed the weighted mean $R_\mathrm{p}/R_*$ values in the common region of the two data sets. We then fitted a horizontal line to the flat spectrum and evaluated our fit using the Bayesian information criterion (BIC) \citep{1978AnSta...6..461S}. This criterion is useful for model selection and penalises complexity, offering a solution to potential over-fitting. We found that the horizontal fit gives a BIC value of 38.0. The generated transmission spectrum of WASP-88b is illustrated in Figure~\ref{fig:transmission_spec}.

On inspection, the spectrum appears featureless with no significant deviations from a straight, horizontal line (all irregularities are within 2$\sigma$ of this mark). This rather flat shape is enhanced by the complete absence of the sodium and potassium features in the wavelength regions of $\sim$5890 and $\sim$7700 \AA, respectively. Another characteristic of the spectrum is the steady upward slope towards shorter wavelengths that may indicate possible scattering in the atmosphere from small particles. We investigated the observed slope following the reasoning of \cite{2008A&A...481L..83L}, where atmospheric opacity as a function of wavelength is described by a scattering cross-section in the form $\sigma=\sigma_{0} ({\lambda} / {\lambda_{0}})^{\gamma}$. Here, index $\gamma$ defines the scattering slope, which can be determined from the transmission spectrum as it is proportional to the slope $\frac{\mathrm{d} \left(R_\mathrm{p} / R_{*}\right)}{\mathrm{~d} \ln \lambda}$:
\begin{equation}
    \gamma=\frac{R_{*}}{H} \frac{\mathrm{d} \left( R_{\mathrm{p}}/ R_{*} \right)}{\mathrm{~d} \ln \lambda}.
\end{equation}
In this equation, $H$ is the atmospheric scale height and is given by $\frac{k_\mathrm{B} T_\mathrm{eq}}{\mu g_\mathrm{p}}$, where $k_\mathrm{B}$ is the Boltzmann constant, $T_\mathrm{eq}$ is the equilibrium temperature, $\mu$ is the mean molecular weight and $g_\mathrm{p}$ is the surface gravity of the planet. Fitting a line with two free parameters (i.e. slope and intercept) to the entire spectrum results in a BIC value of 25.8, which is lower than the value found from the horizontal line fit and indicates that the scattering fit is better ($\Delta\mathrm{BIC}=12.2$). From the slope and the values of stellar radius, planet gravity and planet equilibrium temperature from Table~\ref{tab:tess} we estimated an index value of $\gamma=-12.3^{+2.7}_{-3.0}$. This value suggests greatly enhanced scattering and cannot stem from Rayleigh scattering alone ($\gamma=-4$). Nevertheless, excess scattering is not unusual among hot Jupiters and has been observed before \citep[e.g.][]{2020AJ....160...51A,2021MNRAS.500.5420C}.

\section{Discussion}
\label{sec:discussion}

In this section we set our transmission spectrum up against simulated atmospheres in an attempt to explain the observed slope and featureless shape. We examined the data using both forward models and retrieval techniques and evaluated our findings with respect to theoretical predictions and currently observed trends in exoplanets with similar characteristics.

\subsection{Generic Grid}
\label{sec:goyal_forward_models}
We first compared the observed transmission spectrum to synthetic spectra from a set of distinct model atmospheres. \citet{2018MNRAS.474.5158G} created an extensive grid of forward models based on the one-dimensional plane-parallel radiative-convective equilibrium ATMO model \citep{2014A&A...564A..59A,2015ApJ...804L..17T,2016ApJ...817L..19T}. The library is updated on a regular basis \citep{2019MNRAS.482.4503G,2019MNRAS.486..783G,2020MNRAS.498.4680G} and has been used previously to decipher exoplanet atmospheres \citep[e.g][]{2020MNRAS.494.5449C,2020MNRAS.497.5155W}. We used the publicly available generic version\footnote{\texttt{https://exoctk.stsci.edu/generic}} that assumes isothermal pressure-temperature profiles, equilibrium chemistry and includes opacities from 19 chemical species and from collision induced absorption due to H$_2$-H$_2$ and H$_2$-He interactions \citep{2019MNRAS.482.4503G}. Model atmospheres with varying temperature, surface gravity, metallicity, C/O ratio, scattering hazes, uniform clouds and condensation scheme were considered. An in-depth description of the methodology, the code and the grid parameter setup can be found in \citet{2019MNRAS.482.4503G}.

For our purposes, we adopted a surface gravity of 5 m s$^{-2}$, which is close to the computed value of 4.84 m s$^{-2}$, and considered a reasonable range of planetary temperatures (800 to 2000\,K) in steps of 100\,K. We also made a further assumption that the structure of the atmosphere follows a solar abundance, corresponding to a solar C/O ratio, and that the system is described by a solar metallicity \citep[in agreement with][]{2014A&A...563A.143D,2017A&A...600A..69A}. Since condensation is computed in two different ways (locally only or with rainout), we explored both cases. In the local condensation approach, each atmospheric layer is independent and any material that forms condensates is depleted only from that specific layer. On the other hand, in the rainout scenario, condensing material is depleted from the local layer of the atmosphere and all layers above it under the assumption that any droplets created will sink to deeper layers leaving the low-pressure, upper atmospheric layers devoid of condensate species. Finally, we examined three parametrisations: a clear atmosphere, an atmosphere with a uniform cloud deck described by a cloudiness factor of 1, and an atmosphere with enhanced Rayleigh scattering expressed by a wavelength-dependent haze enhancement factor of 1100 \citep[see][for additional details]{2019MNRAS.482.4503G}.

To determine the best fit model, we computed the mean value of the model transit depth for each respective bin and performed a least squares minimisation by considering the vertical displacement between the measured and computed values of $R_p/R_*$ as the only free parameter. The best model is then established from the lowest BIC value. We noticed that the choice of condensation scheme had no significant effect on the simulated transmission spectrum but we opted to focus our investigation on the rainout scenario for easier comparisons with the models presented in Section~\ref{sec:platon_forward_models}. We find that the model transmission spectrum resembling a hazy atmosphere at 1800\,K results in the best match to our measured data, with the cloud-free and cloudy cases performing slightly worse, obtaining $\Delta$BIC values of 6.4 and 6.6, respectively, when the same temperature is assumed.

\begin{figure}
\centering
\includegraphics[width=0.5\textwidth,height=0.5\textheight,keepaspectratio]{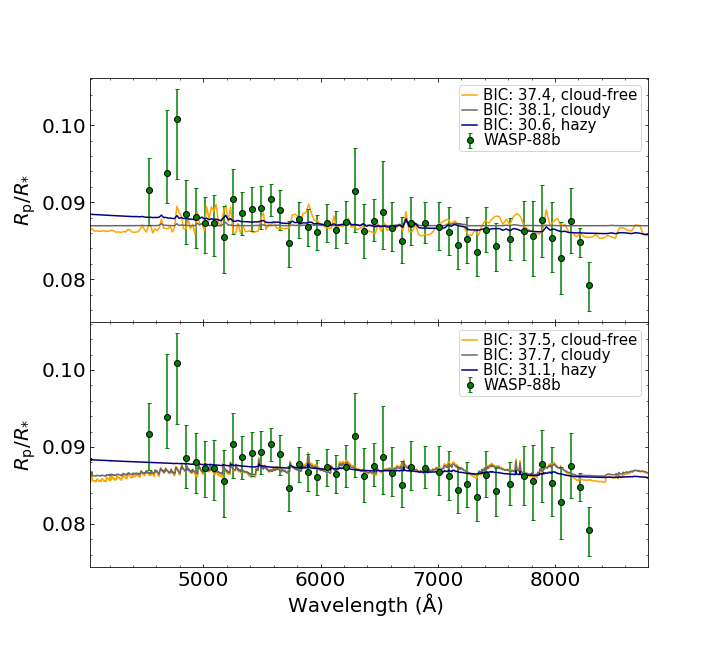}
\caption{The transmission spectrum compared to forward models using \textsc{platon} \citep[top,][]{2019PASP..131c4501Z} and the generic grid \citep[bottom,][]{2019MNRAS.482.4503G}. We fit for three distinct simulated atmospheres (clear, cloudy and hazy) and find a marginal preference for the hazy case, even though all cases produce relatively good fits.}
\label{fig:forward_models}
\end{figure}

\subsection{PLATON}
\label{sec:platon_forward_models}
Another, more flexible, tool that allows comparisons with theoretical spectra obtained through forward modelling is \textsc{platon} \citep[PLanetary Atmospheric Tool for Observer Noobs,][]{2019PASP..131c4501Z,2020ApJ...899...27Z}. This Python package offers a somewhat wider range of temperatures, metallicities and C/O ratios while making most of the same assumptions (isothermal pressure-temperature profiles, equilibrium chemistry). It also provides a choice between an atmosphere where condensation occurs (corresponding to the rainout mode in the generic grid) and one where all chemical species remain in their gas-phase. Additionally, it includes a larger set of opacities (from over 30 chemical species), although it should be pointed out that most of the added molecules have null impact on the generated transmission spectrum and, thus, any disparities from the grid of \citet{2019MNRAS.482.4503G} are primarily attributed to differences in the opacity tables of the most prominent chemical species. One key distinction from \citet{2019MNRAS.482.4503G} is that cloudiness here is described by a cloud-top pressure (i.e. an atmospheric layer where clouds are formed and below which no light can penetrate) instead of a cloudiness factor. Furthermore, the wide supported range in the cloud-top pressure and the scattering factor make \textsc{platon} an ideal tool to explore extremities in the form of very high altitude clouds or super-Rayleigh scattering.

Here, we utilize \textsc{platon} as an extra probe into the gaseous envelope of WASP-88b. To produce a relatively accurate theoretical model, we initialise the planet mass and the planet radius to the values reported in Table~\ref{tab:tess}. We do the same for the star radius while we take the values for the star's effective temperature and metallicity from \citet{2017A&A...600A..69A} and assume a solar C/O ratio. The model transmission spectrum is then generated from the grid through interpolation and we follow the same fitting process using the same range of planetary temperatures described in Section~\ref{sec:goyal_forward_models}. For comparability reasons, we adopt the rainout condensation scheme, although we recognise that the initial parameter configuration is slightly different in this instance. This is because our main goal is to understand the atmosphere of WASP-88b and not to evaluate the performance of various libraries \citep[see][for a direct comparison between the two libraries]{2019PASP..131c4501Z}. Furthermore, it should be noted that due to the different way the cloud deck is portrayed in this case, we expect some deviation from the previous result. After setting the scattering factor to 1100 to simulate haze and the cloud-top pressure to 1 Pa to model a high-altitude cloud deck, we find that our best-fit transmission spectrum advocates a temperature of 1400K and the presence of haze in the atmosphere of WASP-88b, whereas clouds and a clear atmosphere are less likely ($\Delta$BIC values of 7.5 and 6.8 at this temperature). The results here reveal a lower temperature at the day-night terminator region but are very similar to the outcome in Section~\ref{sec:goyal_forward_models} and verify our initial assessment of an upward Rayleigh slope towards the bluer wavelengths (see Section~\ref{sec:transmission_spectrum}) in an otherwise flat transmission spectrum.

Figure~\ref{fig:forward_models} shows the produced transmission spectrum, along with the best-fit forward models from the generic grid and \textsc{platon}, following our main methodology described in Sections~\ref{sec:analysis} and~\ref{sec:transmission_spectrum}. A cloudy atmosphere for WASP-88b was also inferred from parametric light curve fits. These fits included various polynomial combinations of external systematic parameters without cross terms and up to a second degree. However, we found that the parametric approach is likely affected by unaccounted-for systematics and leads to results where the models are unable to fit the data properly.

For completeness, we also employed \textsc{platon} to check how stellar activity can influence the transmission spectrum. We examined a range of spot temperatures and fractional coverages using forward models. We find a slightly better fit to the transmission spectrum but only for implausibly large spot coverage fractions. We therefore do not take this as evidence for the presence of stellar activity in WASP-88.

\subsection{AURA}
\label{sec:aura_retrievals}

\begin{figure*}

\includegraphics[width=\linewidth]{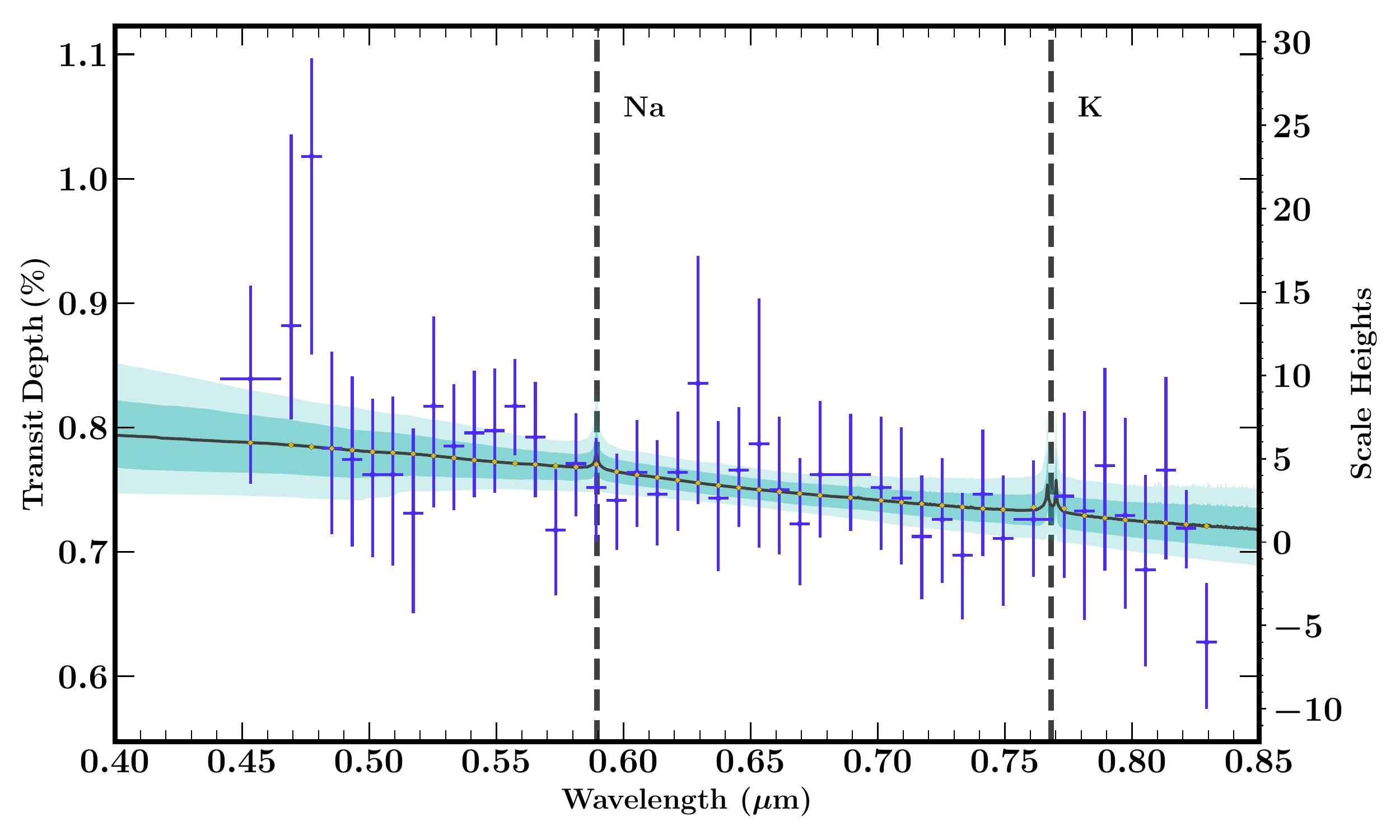}
\caption{The median retrieved transmission spectrum of our \textsc{aura} retrieval (black) and corresponding 1- and 2-$\sigma$ contours (dark and light turquoise, respectively. Vertical dashed lines denote the locations of the Na and K absorption peaks.}
\label{fig:aura_retrieved_spectrum}
\end{figure*}

We lastly analysed the observed transmission spectrum using the latest version of the AURA free retrieval code \citep[]{Pinhas2018, Welbanks2019a}, which couples a forward model-generating component with the PyMultiNest package for parameter estimation and Bayesian model comparison \citep{Buchner2014}. \textsc{aura} generates forward models using a line-by-line radiative transfer calculation, treating the terminator atmosphere as plane-parallel and in hydrostatic equilibrium while assuming that chemical species are uniformly distributed in altitude. 

In this work, we consider atmospheric opacity contributions from H$_2$O \citep{Rothman2010}, Na and K, with the latter two including the effects of H$_2$ broadening \citep{Welbanks2019b}. Opacities arising from H$_2$-H$_2$ and H$_2$-He collision-induced absorption \citep{Richard2012} are also considered. We additionally include the effects of clouds and scattering hazes, modelling clouds as grey opacity and incorporating scattering hazes as a modification to Rayleigh scattering above the cloud deck, given by: $\sigma = a \sigma_0 ({\lambda} / {\lambda_0})^\gamma$, where $\sigma_0 = 5.31 \times 10^{-27}$ cm$^2$ is the H$_2$ Rayleigh scattering cross-section at $\lambda_0 = 350$~nm, while $a$ and $\gamma$ are free parameters. Lastly, we parametrise the terminator pressure-temperature profile using the six-parameter prescription of \citet{Madhusudhan2009}. Our model atmosphere has 14 free parameters in total: three for the volume mixing ratios of H$_2$O, Na and K, six for the pressure-temperature profile, four for clouds, hazes and their fractional coverage and one for the reference pressure at the planet's radius.

For the volume mixing ratios of the three chemical species we use log-uniform priors ranging from 10$^{-12}$ up to 10$^{-1}$. For the temperature at the top of the atmosphere, we use a uniform prior between 800 and 2000K to allow for a broad range of pressure-temperature profiles while excluding any unphysical solutions. For the two haze parameters, we use priors that are log-uniform between 10$^{-4}$- 10$^{10}$ for $a$ and uniform between $-20$ and 2 for $\gamma$.

\begin{figure*}
\centering
\includegraphics[width=\textwidth]{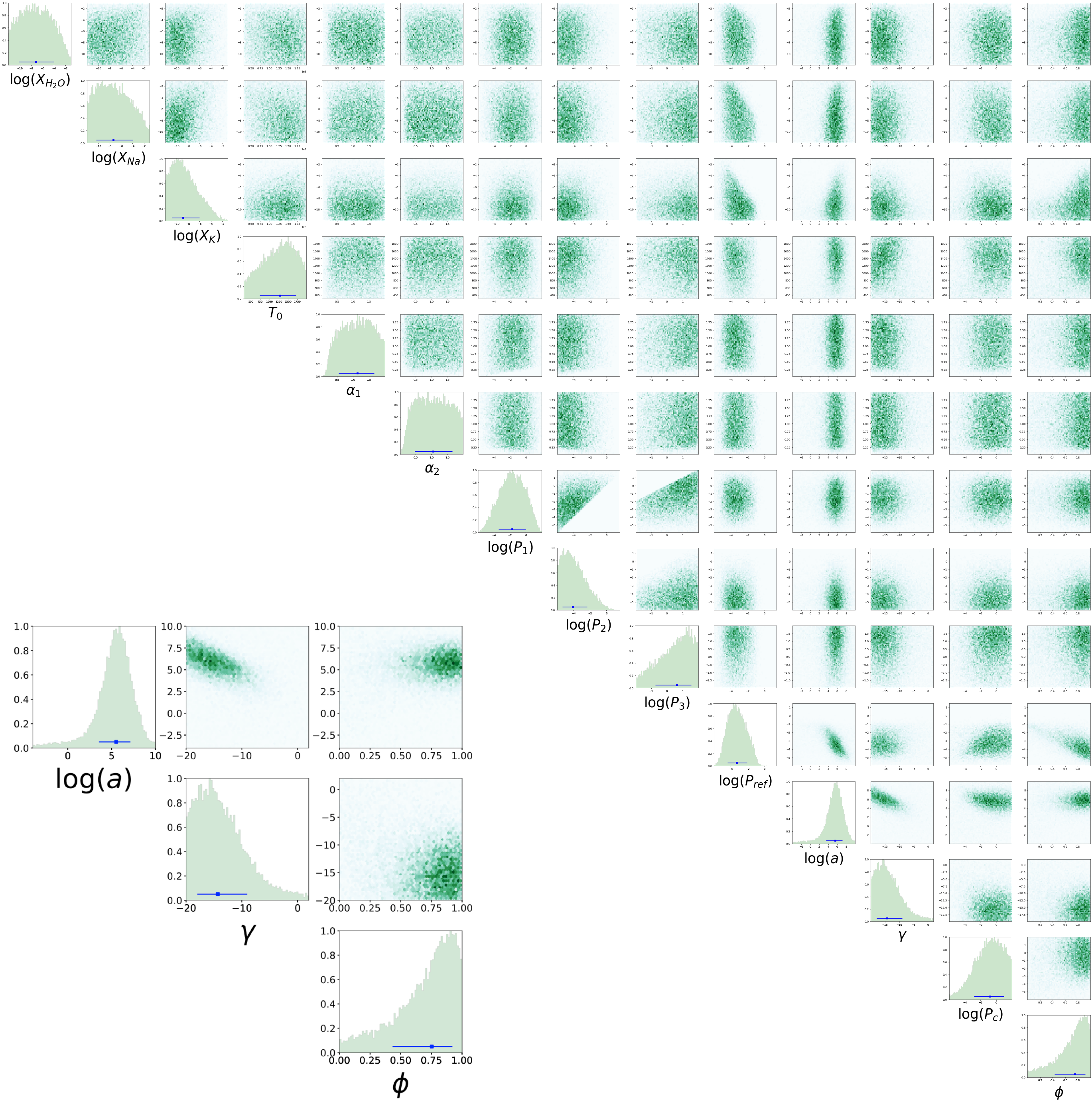}
\caption{Marginalised posterior probability distributions obtained from our \textsc{aura} free retrieval. While the mixing ratios of the three chemical species remain unconstrained, we are able to constrain the two haze parameters, $a$ and $\gamma$. We show the posterior distributions of the two haze parameters, as well as that of the fractional coverage of clouds and hazes, $\phi$, at the bottom left for clarity.}
\label{fig:aura_posterior}
\end{figure*}

Our retrievals are unable to constrain the abundances of Na, K or H$_2$O, finding that the data are best explained by an effectively featureless transmission spectrum with a slight slope due to the presence of scattering from high-altitude hazes. The median transmission spectrum as well as the 1- and 2-$\sigma$ contours are shown in Figure \ref{fig:aura_retrieved_spectrum}. We constrain the two haze parameters to be $\mathrm{log} (a) = 5.53^{+1.56}_{-1.91}$ and $\gamma = -14.42^{+4.88}_{-3.64}$. The full marginalised posterior probability distributions are shown in Figure \ref{fig:aura_posterior}. To quantify the significance of our haze detection using Bayesian model comparison, we carried out a second retrieval using an atmospheric model including clouds but not hazes, fixing the scattering parameters to values corresponding to simple H$_2$ Rayleigh scattering, i.e.\ $a = 1$ and $\gamma = -4$. This retrieval used clouds to explain the data, yielding a flat spectrum. We find that the detection significance for scattering hazes is marginal, at 2.5 $\sigma$. This indicates that while high-altitude scattering hazes are most likely present, we are unable to definitively rule out clouds as the cause for the flat spectrum.

We do not find any significant evidence for the presence of any other chemical species in the atmosphere. For the case described above, the abundances of H$_2$O, Na and K are largely unconstrained, as shown in Figure~\ref{fig:aura_posterior}, with 2-$\sigma$ upper-limits approaching mixing ratios of $\sim$10$^{-2}$ for H$_2$O and Na and $\sim$10$^{-4}$ for K. We also investigated the possibility of other chemical species by conducting retrievals involving various high-temperature molecules including TiO, VO, AlO as well as several metal hydrides, but found no significant evidence for any of these. Finally, we also investigated the possibility of stellar activity impacting observations. We conducted retrievals using the AURA retrieval code's stellar heterogeneity functionality, which models the effects of starspots and faculae, as described in \citet{Pinhas2018} and was recently used to analyse the spectrum of WASP-110~b \citep{Nikolov2021}. Our retrievals found no evidence for the impact of stellar heterogeneity on the transmission spectrum.

\subsection{WASP-88b in context}
\label{sec:context}

The indication that WASP-88b is likely to be hazy and/or cloudy is not very surprising. Widespread clouds have so far been found almost ubiquitously among hot Jupiters probed in transmission and were predicted by \citet{2005MNRAS.364..649F}. Such findings have inspired several efforts to quantify the expected cloud coverage for various planetary atmospheres \citep{2016Natur.529...59S,2016ApJ...817L..16S,2016ApJ...826L..16H}. All these studies use metrics based on the strength and width of certain spectral lines and suggest that the fraction of clouds and hazes at high altitudes is inversely proportional to the level of irradiation. In other words, planets with lower equilibrium temperatures tend to have cloudier atmospheres, which translates to muted or even blocked atomic/molecular features. Moreover, \citet{2016ApJ...817L..16S} also defined a surface gravity threshold above which planets are more likely to be cloud-free. According to this study, planets with an equilibrium temperature below 700K or a log surface gravity below 2.8 have a higher chance to be cloudier. Interestingly, WASP-88b ($T_{\rm eq}=1737~K$, log~$g_{\rm p}=2.68$) meets the second criterion but not the first one. This should come as no surprise, however, as there is now mounting evidence that clouds can be common even at higher equilibrium temperatures \citep[e.g.][]{2017ApJ...835L..12W,2019MNRAS.482.2065E,2020AJ....160...51A,2020A&A...642A..98Y,2020AJ....160..230M}.

A few notable examples of planets with similar equilibrium temperatures and relatively low surface gravities, for which a transmission spectrum in the entire optical regime has been obtained, include HAT-P-32Ab \citep{2020AJ....160...51A}, HAT-P-41b \citep{2020AJ....159..204W,2021AJ....161...51S}, WASP-17b \citep{2019MNRAS.482.1485P} and WASP-31b \citep{2020AJ....160..230M}. Most planets in this sample were found to have some level of cloudiness and/or haziness, with WASP-17b showing evidence of K \citep{2016A&A...596A..47S} and Na \citep{2011MNRAS.412.2376W,2012MNRAS.426.2483Z,2016Natur.529...59S} absorption. Signs of K absorption were initially detected for WASP-31b as well \citep{2015MNRAS.446.2428S}, but later ground-based observations at low- and high-resolution confidently disproved this space-based result, finding no trace of the alkali metal in the atmosphere of this planet \citep{2017MNRAS.467.4591G,2019MNRAS.482..606G,2020AJ....160..230M}. Furthermore, a slope in the visible region due to scattering in the atmosphere can also be found in HAT-P-32Ab \citep{2016A&A...590A.100M,2020AJ....160...51A} and WASP-31b \citep{2020AJ....160..230M}. It therefore seems that WASP-88b's characteristics are not unique among this small group of planets, although it should be added that the scattering slope in this case is more than one and a half times steeper than the ones found for HAT-P-32Ab and WASP-31b.

Such a steep slope is quite challenging in its interpretation and a heterogeneous stellar surface due to increased magnetic activity could be a plausible reason for this. However, our photometric variability analysis from WASP and TESS and additional spectroscopic data that probe the chromosphere show no signs of magnetic activity (see Section~\ref{sec:the_system}). Furthermore, a retrieval analysis with AURA confirms that starspots and faculae do not have an impact on the transmission spectrum. This strongly indicates that the transmission spectrum is not affected by stellar phenomena and that the slope is best explained by physical mechanisms within the planetary atmosphere.

Given the planet's equilibrium temperature, it is possible that the atmospheric conditions favour the condensation of silicate species, although the loosely-constrained, lower temperature value obtained from the retrievals suggests that sulphide condensates could be dominant at the day-night terminator \citep{2016ApJ...828...22P}. The same study also infers a lack of corundum and iron clouds at these temperatures. \citet{2017MNRAS.471.4355P} go a step further and show that a steep slope, such as the one observed for WASP-88b, could be a signature of sulphide clouds. However, a more recent study by \citet{2020NatAs...4..951G} disputes the idea of metal sulphide cloud formation due to nucleation energy barriers. Another potential explanation for the super-Rayleigh slope seen here could be the formation of photochemical haze in an atmosphere with very efficient eddy mixing \citep{2020ApJ...895L..47O}, but WASP-88b falls somewhat outside the reported equilibrium temperature range for this process (1000 to 1500\,K). Future observations with higher precision will be crucial for the understanding of the atmosphere of this inflated hot Jupiter.

\section{Conclusion}
\label{sec:conclusion}

We present the first transmission spectrum of WASP-88b: a hot, gaseous, transiting planet with a low density and a low surface gravity. A revision of the system parameters from TESS is mostly in line with the reported parameters in the discovery paper, with some minor adjustments to the stellar and planetary radii leading to a slightly higher planet-to-star radius ratio. This higher value is backed up from low-resolution spectroscopy with the ground-based VLT FORS2 spectrograph. We employed this instrument to observe two complete transits of WASP-88b using two different gratings that explore the optical regime and cover the wavelength region between 4413 and 8333 \AA. We then constructed a combined transmission spectrum from a total of 45 transit depth values. We found that the spectrum has an overall featureless shape and an enhanced upward slope towards shorter wavelengths. Subsequent analysis with atmospheric forward models and retrievals unveiled the plausible presence of high altitude haze and did not eliminate the possibility of clouds.

A vital check when spectral slopes of this magnitude are observed is to examine whether stellar activity plays any role in this. However, we find no signs of photometric variability and/or chromospheric emission and our retrievals with AURA do not detect evidence of any external influence from the star on the shape of the transmission spectrum. We consider this information more than enough to conclude with high confidence that the star is inactive and, therefore, that stellar activity is likely not responsible for the slope seen in the transmission spectrum of WASP-88b.

Nevertheless, additional monitoring is required to get a more precise picture of the atmosphere and to decisively rule out a clear atmosphere. Complementary observations with the \textit{HST} and the \textit{JWST} in the near-infrared are important to further constrain the atmospheric properties of this planet. The presence or absence of H$_2$O features in the transmission spectrum could help distinguish between the cloudy and hazy possibilities, so a water abundance estimation could provide an idea of how cloudy this atmosphere really is.

\section*{Acknowledgements}
The authors are grateful to the reviewer for their constructive comments on the manuscript. This work is based on observations collected at the European Organization for Astronomical Research in the Southern Hemisphere under European Southern Observatory programme 199.C-0467(D). PS is supported by a UK Science and Technology Facilities Council (STFC) studentship. Ch.H. acknowledges funding from the European Union H2020-MSCA-ITN-2019 under Grant Agreement no. 860470 (CHAMELEON). NM acknowledges funding from the UKRI Future Leaders Scheme (MR/T040866/1), Science and Technology Facilities Council Consolidated Grant (ST/R000395/1) and Leverhulme Trust research project grant (RPG-2020-82).

\section*{Data Availability}

The VLT FORS2 data are publicly available on the ESO archive under programme 199.C-0467(D). The data collected by the TESS mission are publicly available from the Mikulski Archive for Space Telescopes (MAST) at the Space Telescope Science Institure (STScI). Funding for the TESS mission is provided by the NASA's Science Mission Directorate. STScI is operated by the Association of Universities for Research in Astronomy, Inc., under NASA contract NAS5-26555. Support for MAST for non-HST data is provided by the NASA Office of Space Science via grant NNX13AC07G and by other grants and contracts. 



\bibliographystyle{mnras}
\bibliography{references}

\begin{thebibliography}{}
\makeatletter
\relax
\def\mn@urlcharsother{\let\do\@makeother \do\$\do\&\do\#\do\^\do\_\do\%\do\~}
\def\mn@doi{\begingroup\mn@urlcharsother \@ifnextchar [ {\mn@doi@}
  {\mn@doi@[]}}
\def\mn@doi@[#1]#2{\def\@tempa{#1}\ifx\@tempa\@empty \href
  {http://dx.doi.org/#2} {doi:#2}\else \href {http://dx.doi.org/#2} {#1}\fi
  \endgroup}
\def\mn@eprint#1#2{\mn@eprint@#1:#2::\@nil}
\def\mn@eprint@arXiv#1{\href {http://arxiv.org/abs/#1} {{\tt arXiv:#1}}}
\def\mn@eprint@dblp#1{\href {http://dblp.uni-trier.de/rec/bibtex/#1.xml}
  {dblp:#1}}
\def\mn@eprint@#1:#2:#3:#4\@nil{\def\@tempa {#1}\def\@tempb {#2}\def\@tempc
  {#3}\ifx \@tempc \@empty \let \@tempc \@tempb \let \@tempb \@tempa \fi \ifx
  \@tempb \@empty \def\@tempb {arXiv}\fi \@ifundefined
  {mn@eprint@\@tempb}{\@tempb:\@tempc}{\expandafter \expandafter \csname
  mn@eprint@\@tempb\endcsname \expandafter{\@tempc}}}

\bibitem[\protect\citeauthoryear{{Alam} et~al.,}{{Alam}
  et~al.}{2020}]{2020AJ....160...51A}
{Alam} M.~K.,  et~al., 2020, \mn@doi [\aj] {10.3847/1538-3881/ab96cb}, \href
  {https://ui.adsabs.harvard.edu/abs/2020AJ....160...51A} {160, 51}

\bibitem[\protect\citeauthoryear{{Amundsen}, {Baraffe}, {Tremblin}, {Manners},
  {Hayek}, {Mayne}  \& {Acreman}}{{Amundsen}
  et~al.}{2014}]{2014A&A...564A..59A}
{Amundsen} D.~S.,  {Baraffe} I.,  {Tremblin} P.,  {Manners} J.,  {Hayek} W.,
  {Mayne} N.~J.,   {Acreman} D.~M.,  2014, \mn@doi [\aap]
  {10.1051/0004-6361/201323169}, \href
  {https://ui.adsabs.harvard.edu/abs/2014A&A...564A..59A} {564, A59}

\bibitem[\protect\citeauthoryear{{Andreasen} et~al.,}{{Andreasen}
  et~al.}{2017}]{2017A&A...600A..69A}
{Andreasen} D.~T.,  et~al., 2017, \mn@doi [\aap] {10.1051/0004-6361/201629967},
  \href {https://ui.adsabs.harvard.edu/abs/2017A%26A...600A..69A} {600, A69}

\bibitem[\protect\citeauthoryear{{Appenzeller} et~al.,}{{Appenzeller}
  et~al.}{1998}]{1998Msngr..94....1A}
{Appenzeller} I.,  et~al., 1998, The Messenger, \href
  {https://ui.adsabs.harvard.edu/abs/1998Msngr..94....1A} {94, 1}

\bibitem[\protect\citeauthoryear{{Bean}, {Miller-Ricci Kempton}  \&
  {Homeier}}{{Bean} et~al.}{2010}]{2010Natur.468..669B}
{Bean} J.~L.,  {Miller-Ricci Kempton} E.,   {Homeier} D.,  2010, \mn@doi [\nat]
  {10.1038/nature09596}, \href
  {https://ui.adsabs.harvard.edu/abs/2010Natur.468..669B} {468, 669}

\bibitem[\protect\citeauthoryear{{Bean} et~al.,}{{Bean}
  et~al.}{2011}]{2011ApJ...743...92B}
{Bean} J.~L.,  et~al., 2011, \mn@doi [\apj] {10.1088/0004-637X/743/1/92}, \href
  {https://ui.adsabs.harvard.edu/abs/2011ApJ...743...92B} {743, 92}

\bibitem[\protect\citeauthoryear{{Boffin} et~al.,}{{Boffin}
  et~al.}{2016}]{2016SPIE.9908E..2BB}
{Boffin} H. M.~J.,  et~al., 2016, in {Evans} C.~J.,  {Simard} L.,   {Takami}
  H.,  eds,  Society of Photo-Optical Instrumentation Engineers (SPIE)
  Conference Series Vol. 9908, Ground-based and Airborne Instrumentation for
  Astronomy VI. p. 99082B (\mn@eprint {arXiv} {1607.07237}),
  \mn@doi{10.1117/12.2232094}

\bibitem[\protect\citeauthoryear{{Bohn}, {Southworth}, {Ginski}, {Kenworthy},
  {Maxted}  \& {Evans}}{{Bohn} et~al.}{2020}]{2020A&A...635A..73B}
{Bohn} A.~J.,  {Southworth} J.,  {Ginski} C.,  {Kenworthy} M.~A.,  {Maxted}
  P.~F.~L.,   {Evans} D.~F.,  2020, \mn@doi [\aap]
  {10.1051/0004-6361/201937127}, \href
  {https://ui.adsabs.harvard.edu/abs/2020A&A...635A..73B} {635, A73}

\bibitem[\protect\citeauthoryear{{Brown}}{{Brown}}{2001}]{2001ApJ...553.1006B}
{Brown} T.~M.,  2001, \mn@doi [\apj] {10.1086/320950}, \href
  {https://ui.adsabs.harvard.edu/abs/2001ApJ...553.1006B} {553, 1006}

\bibitem[\protect\citeauthoryear{{Buchner} et~al.,}{{Buchner}
  et~al.}{2014}]{Buchner2014}
{Buchner} J.,  et~al., 2014, \mn@doi [\aap] {10.1051/0004-6361/201322971},
  \href {https://ui.adsabs.harvard.edu/abs/2014A&A...564A.125B} {564, A125}

\bibitem[\protect\citeauthoryear{{Carter} et~al.,}{{Carter}
  et~al.}{2020}]{2020MNRAS.494.5449C}
{Carter} A.~L.,  et~al., 2020, \mn@doi [\mnras] {10.1093/mnras/staa1078}, \href
  {https://ui.adsabs.harvard.edu/abs/2020MNRAS.494.5449C} {494, 5449}

\bibitem[\protect\citeauthoryear{{Charbonneau}, {Brown}, {Noyes}  \&
  {Gilliland}}{{Charbonneau} et~al.}{2002}]{2002ApJ...568..377C}
{Charbonneau} D.,  {Brown} T.~M.,  {Noyes} R.~W.,   {Gilliland} R.~L.,  2002,
  \mn@doi [\apj] {10.1086/338770}, \href
  {https://ui.adsabs.harvard.edu/abs/2002ApJ...568..377C} {568, 377}

\bibitem[\protect\citeauthoryear{{Chen} et~al.,}{{Chen}
  et~al.}{2021}]{2021MNRAS.500.5420C}
{Chen} G.,  et~al., 2021, \mn@doi [\mnras] {10.1093/mnras/staa3555}, \href
  {https://ui.adsabs.harvard.edu/abs/2021MNRAS.500.5420C} {500, 5420}

\bibitem[\protect\citeauthoryear{{Claret}}{{Claret}}{2000}]{2000A&A...363.1081C}
{Claret} A.,  2000, \aap, \href
  {https://ui.adsabs.harvard.edu/abs/2000A%26A...363.1081C} {363, 1081}

\bibitem[\protect\citeauthoryear{{Claret}}{{Claret}}{2017}]{2017A+A...600A..30C}
{Claret} A.,  2017, A\&A, \href {2017A+A...600A..30C} {600, A30}

\bibitem[\protect\citeauthoryear{{Delrez} et~al.,}{{Delrez}
  et~al.}{2014}]{2014A&A...563A.143D}
{Delrez} L.,  et~al., 2014, \mn@doi [\aap] {10.1051/0004-6361/201323204}, \href
  {https://ui.adsabs.harvard.edu/abs/2014A%26A...563A.143D} {563, A143}

\bibitem[\protect\citeauthoryear{{Deming} et~al.,}{{Deming}
  et~al.}{2013}]{2013ApJ...774...95D}
{Deming} D.,  et~al., 2013, \mn@doi [\apj] {10.1088/0004-637X/774/2/95}, \href
  {https://ui.adsabs.harvard.edu/abs/2013ApJ...774...95D} {774, 95}

\bibitem[\protect\citeauthoryear{{Eastman}, {Siverd}  \& {Gaudi}}{{Eastman}
  et~al.}{2010}]{2010PASP..122..935E}
{Eastman} J.,  {Siverd} R.,   {Gaudi} B.~S.,  2010, \mn@doi [\pasp]
  {10.1086/655938}, \href
  {https://ui.adsabs.harvard.edu/abs/2010PASP..122..935E} {122, 935}

\bibitem[\protect\citeauthoryear{{Espinoza} et~al.,}{{Espinoza}
  et~al.}{2019}]{2019MNRAS.482.2065E}
{Espinoza} N.,  et~al., 2019, \mn@doi [\mnras] {10.1093/mnras/sty2691}, \href
  {https://ui.adsabs.harvard.edu/abs/2019MNRAS.482.2065E} {482, 2065}

\bibitem[\protect\citeauthoryear{{Foreman-Mackey}}{{Foreman-Mackey}}{2015}]{2015ascl.soft11015F}
{Foreman-Mackey} D.,  2015, {George: Gaussian Process regression} (\mn@eprint
  {ascl} {1511.015})

\bibitem[\protect\citeauthoryear{{Foreman-Mackey}, {Hogg}, {Lang}  \&
  {Goodman}}{{Foreman-Mackey} et~al.}{2013}]{2013PASP..125..306F}
{Foreman-Mackey} D.,  {Hogg} D.~W.,  {Lang} D.,   {Goodman} J.,  2013, \mn@doi
  [\pasp] {10.1086/670067}, \href
  {https://ui.adsabs.harvard.edu/abs/2013PASP..125..306F} {125, 306}

\bibitem[\protect\citeauthoryear{{Fortney}}{{Fortney}}{2005}]{2005MNRAS.364..649F}
{Fortney} J.~J.,  2005, \mn@doi [\mnras] {10.1111/j.1365-2966.2005.09587.x},
  \href {https://ui.adsabs.harvard.edu/abs/2005MNRAS.364..649F} {364, 649}

\bibitem[\protect\citeauthoryear{{Fortney}, {Lodders}, {Marley}  \&
  {Freedman}}{{Fortney} et~al.}{2008}]{2008ApJ...678.1419F}
{Fortney} J.~J.,  {Lodders} K.,  {Marley} M.~S.,   {Freedman} R.~S.,  2008,
  \mn@doi [\apj] {10.1086/528370}, \href
  {https://ui.adsabs.harvard.edu/abs/2008ApJ...678.1419F} {678, 1419}

\bibitem[\protect\citeauthoryear{{Fortney}, {Shabram}, {Showman}, {Lian},
  {Freedman}, {Marley}  \& {Lewis}}{{Fortney}
  et~al.}{2010}]{2010ApJ...709.1396F}
{Fortney} J.~J.,  {Shabram} M.,  {Showman} A.~P.,  {Lian} Y.,  {Freedman}
  R.~S.,  {Marley} M.~S.,   {Lewis} N.~K.,  2010, \mn@doi [\apj]
  {10.1088/0004-637X/709/2/1396}, \href
  {https://ui.adsabs.harvard.edu/abs/2010ApJ...709.1396F} {709, 1396}

\bibitem[\protect\citeauthoryear{{Gao} et~al.,}{{Gao}
  et~al.}{2020}]{2020NatAs...4..951G}
{Gao} P.,  et~al., 2020, \mn@doi [Nature Astronomy]
  {10.1038/s41550-020-1114-3}, \href
  {https://ui.adsabs.harvard.edu/abs/2020NatAs...4..951G} {4, 951}

\bibitem[\protect\citeauthoryear{{Gardner} et~al.,}{{Gardner}
  et~al.}{2006}]{2006SSRv..123..485G}
{Gardner} J.~P.,  et~al., 2006, \mn@doi [\ssr] {10.1007/s11214-006-8315-7},
  \href {https://ui.adsabs.harvard.edu/abs/2006SSRv..123..485G} {123, 485}

\bibitem[\protect\citeauthoryear{{Gibson}}{{Gibson}}{2014}]{2014MNRAS.445.3401G}
{Gibson} N.~P.,  2014, \mn@doi [\mnras] {10.1093/mnras/stu1975}, \href
  {https://ui.adsabs.harvard.edu/abs/2014MNRAS.445.3401G} {445, 3401}

\bibitem[\protect\citeauthoryear{{Gibson}, {Aigrain}, {Roberts}, {Evans},
  {Osborne}  \& {Pont}}{{Gibson} et~al.}{2012}]{2012MNRAS.419.2683G}
{Gibson} N.~P.,  {Aigrain} S.,  {Roberts} S.,  {Evans} T.~M.,  {Osborne} M.,
  {Pont} F.,  2012, \mn@doi [\mnras] {10.1111/j.1365-2966.2011.19915.x}, \href
  {https://ui.adsabs.harvard.edu/abs/2012MNRAS.419.2683G} {419, 2683}

\bibitem[\protect\citeauthoryear{{Gibson}, {Aigrain}, {Barstow}, {Evans},
  {Fletcher}  \& {Irwin}}{{Gibson} et~al.}{2013a}]{2013MNRAS.428.3680G}
{Gibson} N.~P.,  {Aigrain} S.,  {Barstow} J.~K.,  {Evans} T.~M.,  {Fletcher}
  L.~N.,   {Irwin} P.~G.~J.,  2013a, \mn@doi [\mnras] {10.1093/mnras/sts307},
  \href {https://ui.adsabs.harvard.edu/abs/2013MNRAS.428.3680G} {428, 3680}

\bibitem[\protect\citeauthoryear{{Gibson}, {Aigrain}, {Barstow}, {Evans},
  {Fletcher}  \& {Irwin}}{{Gibson} et~al.}{2013b}]{2013MNRAS.436.2974G}
{Gibson} N.~P.,  {Aigrain} S.,  {Barstow} J.~K.,  {Evans} T.~M.,  {Fletcher}
  L.~N.,   {Irwin} P.~G.~J.,  2013b, \mn@doi [\mnras] {10.1093/mnras/stt1783},
  \href {https://ui.adsabs.harvard.edu/abs/2013MNRAS.436.2974G} {436, 2974}

\bibitem[\protect\citeauthoryear{{Gibson}, {Nikolov}, {Sing}, {Barstow},
  {Evans}, {Kataria}  \& {Wilson}}{{Gibson} et~al.}{2017}]{2017MNRAS.467.4591G}
{Gibson} N.~P.,  {Nikolov} N.,  {Sing} D.~K.,  {Barstow} J.~K.,  {Evans} T.~M.,
   {Kataria} T.,   {Wilson} P.~A.,  2017, \mn@doi [\mnras]
  {10.1093/mnras/stx353}, \href
  {https://ui.adsabs.harvard.edu/abs/2017MNRAS.467.4591G} {467, 4591}

\bibitem[\protect\citeauthoryear{{Gibson}, {de Mooij}, {Evans}, {Merritt},
  {Nikolov}, {Sing}  \& {Watson}}{{Gibson} et~al.}{2019}]{2019MNRAS.482..606G}
{Gibson} N.~P.,  {de Mooij} E. J.~W.,  {Evans} T.~M.,  {Merritt} S.,  {Nikolov}
  N.,  {Sing} D.~K.,   {Watson} C.,  2019, \mn@doi [\mnras]
  {10.1093/mnras/sty2722}, \href
  {https://ui.adsabs.harvard.edu/abs/2019MNRAS.482..606G} {482, 606}

\bibitem[\protect\citeauthoryear{{Goodman} \& {Weare}}{{Goodman} \&
  {Weare}}{2010}]{2010CAMCS...5...65G}
{Goodman} J.,  {Weare} J.,  2010, \mn@doi [Communications in Applied
  Mathematics and Computational Science] {10.2140/camcos.2010.5.65}, \href
  {https://ui.adsabs.harvard.edu/abs/2010CAMCS...5...65G} {5, 65}

\bibitem[\protect\citeauthoryear{{Goyal} et~al.,}{{Goyal}
  et~al.}{2018}]{2018MNRAS.474.5158G}
{Goyal} J.~M.,  et~al., 2018, \mn@doi [\mnras] {10.1093/mnras/stx3015}, \href
  {https://ui.adsabs.harvard.edu/abs/2018MNRAS.474.5158G} {474, 5158}

\bibitem[\protect\citeauthoryear{{Goyal}, {Wakeford}, {Mayne}, {Lewis},
  {Drummond}  \& {Sing}}{{Goyal} et~al.}{2019a}]{2019MNRAS.482.4503G}
{Goyal} J.~M.,  {Wakeford} H.~R.,  {Mayne} N.~J.,  {Lewis} N.~K.,  {Drummond}
  B.,   {Sing} D.~K.,  2019a, \mn@doi [\mnras] {10.1093/mnras/sty3001}, \href
  {https://ui.adsabs.harvard.edu/abs/2019MNRAS.482.4503G} {482, 4503}

\bibitem[\protect\citeauthoryear{{Goyal} et~al.,}{{Goyal}
  et~al.}{2019b}]{2019MNRAS.486..783G}
{Goyal} J.~M.,  et~al., 2019b, \mn@doi [\mnras] {10.1093/mnras/stz755}, \href
  {https://ui.adsabs.harvard.edu/abs/2019MNRAS.486..783G} {486, 783}

\bibitem[\protect\citeauthoryear{{Goyal} et~al.,}{{Goyal}
  et~al.}{2020}]{2020MNRAS.498.4680G}
{Goyal} J.~M.,  et~al., 2020, \mn@doi [\mnras] {10.1093/mnras/staa2300}, \href
  {https://ui.adsabs.harvard.edu/abs/2020MNRAS.498.4680G} {498, 4680}

\bibitem[\protect\citeauthoryear{{Heng}}{{Heng}}{2016}]{2016ApJ...826L..16H}
{Heng} K.,  2016, \mn@doi [\apjl] {10.3847/2041-8205/826/1/L16}, \href
  {https://ui.adsabs.harvard.edu/abs/2016ApJ...826L..16H} {826, L16}

\bibitem[\protect\citeauthoryear{{Hubeny}, {Burrows}  \& {Sudarsky}}{{Hubeny}
  et~al.}{2003}]{2003ApJ...594.1011H}
{Hubeny} I.,  {Burrows} A.,   {Sudarsky} D.,  2003, \mn@doi [\apj]
  {10.1086/377080}, \href
  {https://ui.adsabs.harvard.edu/abs/2003ApJ...594.1011H} {594, 1011}

\bibitem[\protect\citeauthoryear{{Huitson} et~al.,}{{Huitson}
  et~al.}{2013}]{2013MNRAS.434.3252H}
{Huitson} C.~M.,  et~al., 2013, \mn@doi [\mnras] {10.1093/mnras/stt1243}, \href
  {https://ui.adsabs.harvard.edu/abs/2013MNRAS.434.3252H} {434, 3252}

\bibitem[\protect\citeauthoryear{{Jenkins} et~al.,}{{Jenkins}
  et~al.}{2016}]{2016SPIE.9913E..3EJ}
{Jenkins} J.~M.,  et~al., 2016, in Proc.\ SPIE. p. 99133E

\bibitem[\protect\citeauthoryear{{Kanodia} \& {Wright}}{{Kanodia} \&
  {Wright}}{2018}]{2018RNAAS...2a...4K}
{Kanodia} S.,  {Wright} J.,  2018, \mn@doi [Research Notes of the American
  Astronomical Society] {10.3847/2515-5172/aaa4b7}, \href
  {https://ui.adsabs.harvard.edu/abs/2018RNAAS...2a...4K} {2, 4}

\bibitem[\protect\citeauthoryear{{Kirk}, {Wheatley}, {Louden}, {Doyle},
  {Skillen}, {McCormac}, {Irwin}  \& {Karjalainen}}{{Kirk}
  et~al.}{2017}]{2017MNRAS.468.3907K}
{Kirk} J.,  {Wheatley} P.~J.,  {Louden} T.,  {Doyle} A.~P.,  {Skillen} I.,
  {McCormac} J.,  {Irwin} P.~G.~J.,   {Karjalainen} R.,  2017, \mn@doi [\mnras]
  {10.1093/mnras/stx752}, \href
  {https://ui.adsabs.harvard.edu/abs/2017MNRAS.468.3907K} {468, 3907}

\bibitem[\protect\citeauthoryear{{Kopal}}{{Kopal}}{1950}]{1950HarCi.454....1K}
{Kopal} Z.,  1950, Harvard College Observatory Circular, \href
  {https://ui.adsabs.harvard.edu/abs/1950HarCi.454....1K} {454, 1}

\bibitem[\protect\citeauthoryear{{Kreidberg}}{{Kreidberg}}{2015}]{2015PASP..127.1161K}
{Kreidberg} L.,  2015, \mn@doi [\pasp] {10.1086/683602}, \href
  {https://ui.adsabs.harvard.edu/abs/2015PASP..127.1161K} {127, 1161}

\bibitem[\protect\citeauthoryear{{Lecavelier Des Etangs}, {Pont},
  {Vidal-Madjar}  \& {Sing}}{{Lecavelier Des Etangs}
  et~al.}{2008}]{2008A&A...481L..83L}
{Lecavelier Des Etangs} A.,  {Pont} F.,  {Vidal-Madjar} A.,   {Sing} D.,  2008,
  \mn@doi [\aap] {10.1051/0004-6361:200809388}, \href
  {https://ui.adsabs.harvard.edu/abs/2008A&A...481L..83L} {481, L83}

\bibitem[\protect\citeauthoryear{{Lendl} et~al.,}{{Lendl}
  et~al.}{2016}]{2016A&A...587A..67L}
{Lendl} M.,  et~al., 2016, \mn@doi [\aap] {10.1051/0004-6361/201527594}, \href
  {https://ui.adsabs.harvard.edu/abs/2016A%26A...587A..67L} {587, A67}

\bibitem[\protect\citeauthoryear{{Lendl}, {Cubillos}, {Hagelberg},
  {M{\"u}ller}, {Juvan}  \& {Fossati}}{{Lendl}
  et~al.}{2017}]{2017A&A...606A..18L}
{Lendl} M.,  {Cubillos} P.~E.,  {Hagelberg} J.,  {M{\"u}ller} A.,  {Juvan} I.,
   {Fossati} L.,  2017, \mn@doi [\aap] {10.1051/0004-6361/201731242}, \href
  {https://ui.adsabs.harvard.edu/abs/2017A%26A...606A..18L} {606, A18}

\bibitem[\protect\citeauthoryear{{Madhusudhan} \& {Seager}}{{Madhusudhan} \&
  {Seager}}{2009}]{Madhusudhan2009}
{Madhusudhan} N.,  {Seager} S.,  2009, \mn@doi [\apj]
  {10.1088/0004-637X/707/1/24}, \href
  {https://ui.adsabs.harvard.edu/abs/2009ApJ...707...24M} {707, 24}

\bibitem[\protect\citeauthoryear{{Magic}, {Chiavassa}, {Collet}  \&
  {Asplund}}{{Magic} et~al.}{2015}]{2015A&A...573A..90M}
{Magic} Z.,  {Chiavassa} A.,  {Collet} R.,   {Asplund} M.,  2015, \mn@doi
  [\aap] {10.1051/0004-6361/201423804}, \href
  {https://ui.adsabs.harvard.edu/abs/2015A%26A...573A..90M} {573, A90}

\bibitem[\protect\citeauthoryear{{Mallonn} \& {Strassmeier}}{{Mallonn} \&
  {Strassmeier}}{2016}]{2016A&A...590A.100M}
{Mallonn} M.,  {Strassmeier} K.~G.,  2016, \mn@doi [\aap]
  {10.1051/0004-6361/201527898}, \href
  {https://ui.adsabs.harvard.edu/abs/2016A%26A...590A.100M} {590, A100}

\bibitem[\protect\citeauthoryear{{Mancini} et~al.,}{{Mancini}
  et~al.}{2013}]{2013MNRAS.436....2M}
{Mancini} L.,  et~al., 2013, MNRAS, \href {2013MNRAS.436....2M} {436, 2}

\bibitem[\protect\citeauthoryear{{Mandel} \& {Agol}}{{Mandel} \&
  {Agol}}{2002}]{2002ApJ...580L.171M}
{Mandel} K.,  {Agol} E.,  2002, \mn@doi [\apjl] {10.1086/345520}, \href
  {https://ui.adsabs.harvard.edu/abs/2002ApJ...580L.171M} {580, L171}

\bibitem[\protect\citeauthoryear{{Maxted} et~al.,}{{Maxted}
  et~al.}{2020}]{2020MNRAS.498..332M}
{Maxted} P.~F.~L.,  et~al., 2020, MNRAS, \href {2020MNRAS.498..332M} {498, 332}

\bibitem[\protect\citeauthoryear{{May}, {Gardner}, {Rauscher}  \&
  {Monnier}}{{May} et~al.}{2020}]{2020AJ....159....7M}
{May} E.~M.,  {Gardner} T.,  {Rauscher} E.,   {Monnier} J.~D.,  2020, \mn@doi
  [\aj] {10.3847/1538-3881/ab5361}, \href
  {https://ui.adsabs.harvard.edu/abs/2020AJ....159....7M} {159, 7}

\bibitem[\protect\citeauthoryear{{McCullough}, {Crouzet}, {Deming}  \&
  {Madhusudhan}}{{McCullough} et~al.}{2014}]{2014ApJ...791...55M}
{McCullough} P.~R.,  {Crouzet} N.,  {Deming} D.,   {Madhusudhan} N.,  2014,
  \mn@doi [\apj] {10.1088/0004-637X/791/1/55}, \href
  {https://ui.adsabs.harvard.edu/abs/2014ApJ...791...55M} {791, 55}

\bibitem[\protect\citeauthoryear{{McGruder} et~al.,}{{McGruder}
  et~al.}{2020}]{2020AJ....160..230M}
{McGruder} C.~D.,  et~al., 2020, \mn@doi [\aj] {10.3847/1538-3881/abb806},
  \href {https://ui.adsabs.harvard.edu/abs/2020AJ....160..230M} {160, 230}

\bibitem[\protect\citeauthoryear{{Nikolov}, {Chen}, {Fortney}, {Mancini},
  {Southworth}, {van Boekel}  \& {Henning}}{{Nikolov}
  et~al.}{2013}]{2013A&A...553A..26N}
{Nikolov} N.,  {Chen} G.,  {Fortney} J.~J.,  {Mancini} L.,  {Southworth} J.,
  {van Boekel} R.,   {Henning} T.,  2013, \mn@doi [\aap]
  {10.1051/0004-6361/201321084}, \href
  {https://ui.adsabs.harvard.edu/abs/2013A%26A...553A..26N} {553, A26}

\bibitem[\protect\citeauthoryear{{Nikolov} et~al.,}{{Nikolov}
  et~al.}{2014}]{2014MNRAS.437...46N}
{Nikolov} N.,  et~al., 2014, \mn@doi [\mnras] {10.1093/mnras/stt1859}, \href
  {https://ui.adsabs.harvard.edu/abs/2014MNRAS.437...46N} {437, 46}

\bibitem[\protect\citeauthoryear{{Nikolov}, {Sing}, {Gibson}, {Fortney},
  {Evans}, {Barstow}, {Kataria}  \& {Wilson}}{{Nikolov}
  et~al.}{2016}]{2016ApJ...832..191N}
{Nikolov} N.,  {Sing} D.~K.,  {Gibson} N.~P.,  {Fortney} J.~J.,  {Evans} T.~M.,
   {Barstow} J.~K.,  {Kataria} T.,   {Wilson} P.~A.,  2016, \mn@doi [\apj]
  {10.3847/0004-637X/832/2/191}, \href
  {https://ui.adsabs.harvard.edu/abs/2016ApJ...832..191N} {832, 191}

\bibitem[\protect\citeauthoryear{{Nikolov} et~al.,}{{Nikolov}
  et~al.}{2018}]{2018Natur.557..526N}
{Nikolov} N.,  et~al., 2018, \mn@doi [\nat] {10.1038/s41586-018-0101-7}, \href
  {https://ui.adsabs.harvard.edu/abs/2018Natur.557..526N} {557, 526}

\bibitem[\protect\citeauthoryear{{Nikolov} et~al.,}{{Nikolov}
  et~al.}{2021}]{Nikolov2021}
{Nikolov} N.,  et~al., 2021, arXiv e-prints, \href
  {https://ui.adsabs.harvard.edu/abs/2021arXiv210506522N} {p. arXiv:2105.06522}

\bibitem[\protect\citeauthoryear{{Ohno} \& {Kawashima}}{{Ohno} \&
  {Kawashima}}{2020}]{2020ApJ...895L..47O}
{Ohno} K.,  {Kawashima} Y.,  2020, \mn@doi [\apjl] {10.3847/2041-8213/ab93d7},
  \href {https://ui.adsabs.harvard.edu/abs/2020ApJ...895L..47O} {895, L47}

\bibitem[\protect\citeauthoryear{{Parmentier}, {Fortney}, {Showman}, {Morley}
  \& {Marley}}{{Parmentier} et~al.}{2016}]{2016ApJ...828...22P}
{Parmentier} V.,  {Fortney} J.~J.,  {Showman} A.~P.,  {Morley} C.,   {Marley}
  M.~S.,  2016, \mn@doi [\apj] {10.3847/0004-637X/828/1/22}, \href
  {https://ui.adsabs.harvard.edu/abs/2016ApJ...828...22P} {828, 22}

\bibitem[\protect\citeauthoryear{{Pinhas} \& {Madhusudhan}}{{Pinhas} \&
  {Madhusudhan}}{2017}]{2017MNRAS.471.4355P}
{Pinhas} A.,  {Madhusudhan} N.,  2017, \mn@doi [\mnras]
  {10.1093/mnras/stx1849}, \href
  {https://ui.adsabs.harvard.edu/abs/2017MNRAS.471.4355P} {471, 4355}

\bibitem[\protect\citeauthoryear{{Pinhas}, {Rackham}, {Madhusudhan}  \&
  {Apai}}{{Pinhas} et~al.}{2018}]{Pinhas2018}
{Pinhas} A.,  {Rackham} B.~V.,  {Madhusudhan} N.,   {Apai} D.,  2018, \mn@doi
  [\mnras] {10.1093/mnras/sty2209}, \href
  {https://ui.adsabs.harvard.edu/abs/2018MNRAS.480.5314P} {480, 5314}

\bibitem[\protect\citeauthoryear{{Pinhas}, {Madhusudhan}, {Gandhi}  \&
  {MacDonald}}{{Pinhas} et~al.}{2019}]{2019MNRAS.482.1485P}
{Pinhas} A.,  {Madhusudhan} N.,  {Gandhi} S.,   {MacDonald} R.,  2019, \mn@doi
  [\mnras] {10.1093/mnras/sty2544}, \href
  {https://ui.adsabs.harvard.edu/abs/2019MNRAS.482.1485P} {482, 1485}

\bibitem[\protect\citeauthoryear{{Rackham} et~al.,}{{Rackham}
  et~al.}{2017}]{2017ApJ...834..151R}
{Rackham} B.,  et~al., 2017, \mn@doi [\apj] {10.3847/1538-4357/aa4f6c}, \href
  {https://ui.adsabs.harvard.edu/abs/2017ApJ...834..151R} {834, 151}

\bibitem[\protect\citeauthoryear{Richard et~al.,}{Richard
  et~al.}{2012}]{Richard2012}
Richard C.,  et~al., 2012, \mn@doi [\jqsrt]
  {https://doi.org/10.1016/j.jqsrt.2011.11.004}, 113, 1276

\bibitem[\protect\citeauthoryear{{Ricker} et~al.,}{{Ricker}
  et~al.}{2015}]{2015JATIS...1a4003R}
{Ricker} G.~R.,  et~al., 2015, \mn@doi [Journal of Astronomical Telescopes,
  Instruments, and Systems] {10.1117/1.JATIS.1.1.014003}, \href
  {https://ui.adsabs.harvard.edu/abs/2015JATIS...1a4003R} {1, 014003}

\bibitem[\protect\citeauthoryear{{Roberts}, {Osborne}, {Ebden}, {Reece},
  {Gibson}  \& {Aigrain}}{{Roberts} et~al.}{2012}]{2012RSPTA.37110550R}
{Roberts} S.,  {Osborne} M.,  {Ebden} M.,  {Reece} S.,  {Gibson} N.,
  {Aigrain} S.,  2012, \mn@doi [Philosophical Transactions of the Royal Society
  of London Series A] {10.1098/rsta.2011.0550}, \href
  {https://ui.adsabs.harvard.edu/abs/2012RSPTA.37110550R} {371, 20110550}

\bibitem[\protect\citeauthoryear{Rothman et~al.,}{Rothman
  et~al.}{2010}]{Rothman2010}
Rothman L.,  et~al., 2010, \mn@doi [\jqsrt]
  {https://doi.org/10.1016/j.jqsrt.2010.05.001}, 111, 2139

\bibitem[\protect\citeauthoryear{{Schwarz}}{{Schwarz}}{1978}]{1978AnSta...6..461S}
{Schwarz} G.,  1978, Annals of Statistics, \href
  {https://ui.adsabs.harvard.edu/abs/1978AnSta...6..461S} {6, 461}

\bibitem[\protect\citeauthoryear{{Seager} \& {Sasselov}}{{Seager} \&
  {Sasselov}}{2000}]{2000ApJ...537..916S}
{Seager} S.,  {Sasselov} D.~D.,  2000, \mn@doi [\apj] {10.1086/309088}, \href
  {https://ui.adsabs.harvard.edu/abs/2000ApJ...537..916S} {537, 916}

\bibitem[\protect\citeauthoryear{{Sedaghati}, {Boffin}, {Csizmadia}, {Gibson},
  {Kabath}, {Mallonn}  \& {Van den Ancker}}{{Sedaghati}
  et~al.}{2015}]{2015A&A...576L..11S}
{Sedaghati} E.,  {Boffin} H.~M.~J.,  {Csizmadia} S.,  {Gibson} N.,  {Kabath}
  P.,  {Mallonn} M.,   {Van den Ancker} M.~E.,  2015, \mn@doi [\aap]
  {10.1051/0004-6361/201525822}, \href
  {https://ui.adsabs.harvard.edu/abs/2015A&A...576L..11S} {576, L11}

\bibitem[\protect\citeauthoryear{{Sedaghati} et~al.,}{{Sedaghati}
  et~al.}{2016}]{2016A&A...596A..47S}
{Sedaghati} E.,  et~al., 2016, \mn@doi [\aap] {10.1051/0004-6361/201629090},
  \href {https://ui.adsabs.harvard.edu/abs/2016A%26A...596A..47S} {596, A47}

\bibitem[\protect\citeauthoryear{{Sedaghati}, {Boffin}, {Delrez}, {Gillon},
  {Csizmadia}, {Smith}  \& {Rauer}}{{Sedaghati}
  et~al.}{2017}]{2017MNRAS.468.3123S}
{Sedaghati} E.,  {Boffin} H. M.~J.,  {Delrez} L.,  {Gillon} M.,  {Csizmadia}
  S.,  {Smith} A. M.~S.,   {Rauer} H.,  2017, \mn@doi [\mnras]
  {10.1093/mnras/stx646}, \href
  {https://ui.adsabs.harvard.edu/abs/2017MNRAS.468.3123S} {468, 3123}

\bibitem[\protect\citeauthoryear{{Sheppard} et~al.,}{{Sheppard}
  et~al.}{2021}]{2021AJ....161...51S}
{Sheppard} K.~B.,  et~al., 2021, \mn@doi [\aj] {10.3847/1538-3881/abc8f4},
  \href {https://ui.adsabs.harvard.edu/abs/2021AJ....161...51S} {161, 51}

\bibitem[\protect\citeauthoryear{{Sing}, {Vidal-Madjar}, {D{\'e}sert},
  {Lecavelier des Etangs}  \& {Ballester}}{{Sing}
  et~al.}{2008}]{2008ApJ...686..658S}
{Sing} D.~K.,  {Vidal-Madjar} A.,  {D{\'e}sert} J.~M.,  {Lecavelier des Etangs}
  A.,   {Ballester} G.,  2008, \mn@doi [\apj] {10.1086/590075}, \href
  {https://ui.adsabs.harvard.edu/abs/2008ApJ...686..658S} {686, 658}

\bibitem[\protect\citeauthoryear{{Sing} et~al.,}{{Sing}
  et~al.}{2012}]{2012MNRAS.426.1663S}
{Sing} D.~K.,  et~al., 2012, \mn@doi [\mnras]
  {10.1111/j.1365-2966.2012.21938.x}, \href
  {https://ui.adsabs.harvard.edu/abs/2012MNRAS.426.1663S} {426, 1663}

\bibitem[\protect\citeauthoryear{{Sing} et~al.,}{{Sing}
  et~al.}{2015}]{2015MNRAS.446.2428S}
{Sing} D.~K.,  et~al., 2015, \mn@doi [\mnras] {10.1093/mnras/stu2279}, \href
  {https://ui.adsabs.harvard.edu/abs/2015MNRAS.446.2428S} {446, 2428}

\bibitem[\protect\citeauthoryear{{Sing} et~al.,}{{Sing}
  et~al.}{2016}]{2016Natur.529...59S}
{Sing} D.~K.,  et~al., 2016, \mn@doi [\nat] {10.1038/nature16068}, \href
  {https://ui.adsabs.harvard.edu/abs/2016Natur.529...59S} {529, 59}

\bibitem[\protect\citeauthoryear{{Southworth}}{{Southworth}}{2008}]{2008MNRAS.386.1644S}
{Southworth} J.,  2008, MNRAS, \href {2008MNRAS.386.1644S} {386, 1644}

\bibitem[\protect\citeauthoryear{{Southworth}}{{Southworth}}{2010}]{2010MNRAS.408.1689S}
{Southworth} J.,  2010, MNRAS, \href {2010MNRAS.408.1689S} {408, 1689}

\bibitem[\protect\citeauthoryear{{Southworth}}{{Southworth}}{2012}]{2012MNRAS.426.1291S}
{Southworth} J.,  2012, MNRAS, \href {2012MNRAS.426.1291S} {426, 1291}

\bibitem[\protect\citeauthoryear{{Southworth}}{{Southworth}}{2013}]{2013A+A...557A.119S}
{Southworth} J.,  2013, A\&A, \href {2013A+A...557A.119S} {557, A119}

\bibitem[\protect\citeauthoryear{{Southworth} \& {Evans}}{{Southworth} \&
  {Evans}}{2016}]{2016MNRAS.463...37S}
{Southworth} J.,  {Evans} D.~F.,  2016, \mn@doi [\mnras]
  {10.1093/mnras/stw1943}, \href
  {https://ui.adsabs.harvard.edu/abs/2016MNRAS.463...37S} {463, 37}

\bibitem[\protect\citeauthoryear{{Southworth}, {Bohn}, {Kenworthy}, {Ginski}
  \& {Mancini}}{{Southworth} et~al.}{2020}]{2020A+A...635A..74S}
{Southworth} J.,  {Bohn} A.~J.,  {Kenworthy} M.~A.,  {Ginski} C.,   {Mancini}
  L.,  2020, A\&A, \href {2020A+A...635A..74S} {635, A74}

\bibitem[\protect\citeauthoryear{{Stevenson}}{{Stevenson}}{2016}]{2016ApJ...817L..16S}
{Stevenson} K.~B.,  2016, \mn@doi [\apjl] {10.3847/2041-8205/817/2/L16}, \href
  {https://ui.adsabs.harvard.edu/abs/2016ApJ...817L..16S} {817, L16}

\bibitem[\protect\citeauthoryear{{Sudarsky}, {Burrows}  \& {Pinto}}{{Sudarsky}
  et~al.}{2000}]{sudarsky00}
{Sudarsky} D.,  {Burrows} A.,   {Pinto} P.,  2000, \mn@doi [\apj]
  {10.1086/309160}, \href {http://adsabs.harvard.edu/abs/2000ApJ...538..885S}
  {538, 885}

\bibitem[\protect\citeauthoryear{{Tremblin}, {Amundsen}, {Mourier}, {Baraffe},
  {Chabrier}, {Drummond}, {Homeier}  \& {Venot}}{{Tremblin}
  et~al.}{2015}]{2015ApJ...804L..17T}
{Tremblin} P.,  {Amundsen} D.~S.,  {Mourier} P.,  {Baraffe} I.,  {Chabrier} G.,
   {Drummond} B.,  {Homeier} D.,   {Venot} O.,  2015, \mn@doi [\apjl]
  {10.1088/2041-8205/804/1/L17}, \href
  {https://ui.adsabs.harvard.edu/abs/2015ApJ...804L..17T} {804, L17}

\bibitem[\protect\citeauthoryear{{Tremblin}, {Amundsen}, {Chabrier}, {Baraffe},
  {Drummond}, {Hinkley}, {Mourier}  \& {Venot}}{{Tremblin}
  et~al.}{2016}]{2016ApJ...817L..19T}
{Tremblin} P.,  {Amundsen} D.~S.,  {Chabrier} G.,  {Baraffe} I.,  {Drummond}
  B.,  {Hinkley} S.,  {Mourier} P.,   {Venot} O.,  2016, \mn@doi [\apjl]
  {10.3847/2041-8205/817/2/L19}, \href
  {https://ui.adsabs.harvard.edu/abs/2016ApJ...817L..19T} {817, L19}

\bibitem[\protect\citeauthoryear{{Wakeford} et~al.,}{{Wakeford}
  et~al.}{2013}]{2013MNRAS.435.3481W}
{Wakeford} H.~R.,  et~al., 2013, \mn@doi [\mnras] {10.1093/mnras/stt1536},
  \href {https://ui.adsabs.harvard.edu/abs/2013MNRAS.435.3481W} {435, 3481}

\bibitem[\protect\citeauthoryear{{Wakeford} et~al.,}{{Wakeford}
  et~al.}{2017}]{2017ApJ...835L..12W}
{Wakeford} H.~R.,  et~al., 2017, \mn@doi [\apjl] {10.3847/2041-8213/835/1/L12},
  \href {https://ui.adsabs.harvard.edu/abs/2017ApJ...835L..12W} {835, L12}

\bibitem[\protect\citeauthoryear{{Wakeford} et~al.,}{{Wakeford}
  et~al.}{2020}]{2020AJ....159..204W}
{Wakeford} H.~R.,  et~al., 2020, \mn@doi [\aj] {10.3847/1538-3881/ab7b78},
  \href {https://ui.adsabs.harvard.edu/abs/2020AJ....159..204W} {159, 204}

\bibitem[\protect\citeauthoryear{{Weaver} et~al.,}{{Weaver}
  et~al.}{2020}]{2020AJ....159...13W}
{Weaver} I.~C.,  et~al., 2020, \mn@doi [\aj] {10.3847/1538-3881/ab55da}, \href
  {https://ui.adsabs.harvard.edu/abs/2020AJ....159...13W} {159, 13}

\bibitem[\protect\citeauthoryear{{Welbanks} \& {Madhusudhan}}{{Welbanks} \&
  {Madhusudhan}}{2019}]{Welbanks2019a}
{Welbanks} L.,  {Madhusudhan} N.,  2019, \mn@doi [AJ]
  {10.3847/1538-3881/ab14de}, \href
  {https://ui.adsabs.harvard.edu/abs/2019AJ....157..206W} {157, 206}

\bibitem[\protect\citeauthoryear{{Welbanks}, {Madhusudhan}, {Allard}, {Hubeny},
  {Spiegelman}  \& {Leininger}}{{Welbanks} et~al.}{2019}]{Welbanks2019b}
{Welbanks} L.,  {Madhusudhan} N.,  {Allard} N.~F.,  {Hubeny} I.,  {Spiegelman}
  F.,   {Leininger} T.,  2019, \mn@doi [ApJLett] {10.3847/2041-8213/ab5a89},
  \href {https://ui.adsabs.harvard.edu/abs/2019ApJ...887L..20W} {887, L20}

\bibitem[\protect\citeauthoryear{{Wilson} et~al.,}{{Wilson}
  et~al.}{2020}]{2020MNRAS.497.5155W}
{Wilson} J.,  et~al., 2020, \mn@doi [\mnras] {10.1093/mnras/staa2307}, \href
  {https://ui.adsabs.harvard.edu/abs/2020MNRAS.497.5155W} {497, 5155}

\bibitem[\protect\citeauthoryear{{Winn}}{{Winn}}{2010}]{2010arXiv1001.2010W}
{Winn} J.~N.,  2010, arXiv e-prints, \href
  {https://ui.adsabs.harvard.edu/abs/2010arXiv1001.2010W} {p. arXiv:1001.2010}

\bibitem[\protect\citeauthoryear{{Wood}, {Maxted}, {Smalley}  \& {Iro}}{{Wood}
  et~al.}{2011}]{2011MNRAS.412.2376W}
{Wood} P.~L.,  {Maxted} P.~F.~L.,  {Smalley} B.,   {Iro} N.,  2011, \mn@doi
  [\mnras] {10.1111/j.1365-2966.2010.18061.x}, \href
  {https://ui.adsabs.harvard.edu/abs/2011MNRAS.412.2376W} {412, 2376}

\bibitem[\protect\citeauthoryear{{Yan} et~al.,}{{Yan}
  et~al.}{2020}]{2020A&A...642A..98Y}
{Yan} F.,  et~al., 2020, \mn@doi [\aap] {10.1051/0004-6361/201937265}, \href
  {https://ui.adsabs.harvard.edu/abs/2020A&A...642A..98Y} {642, A98}

\bibitem[\protect\citeauthoryear{{Zhang}, {Chachan}, {Kempton}  \&
  {Knutson}}{{Zhang} et~al.}{2019}]{2019PASP..131c4501Z}
{Zhang} M.,  {Chachan} Y.,  {Kempton} E. M.~R.,   {Knutson} H.~A.,  2019,
  \mn@doi [\pasp] {10.1088/1538-3873/aaf5ad}, \href
  {https://ui.adsabs.harvard.edu/abs/2019PASP..131c4501Z} {131, 034501}

\bibitem[\protect\citeauthoryear{{Zhang}, {Chachan}, {Kempton}, {Knutson}  \&
  {Chang}}{{Zhang} et~al.}{2020}]{2020ApJ...899...27Z}
{Zhang} M.,  {Chachan} Y.,  {Kempton} E. M.~R.,  {Knutson} H.~A.,   {Chang}
  W.~H.,  2020, \mn@doi [\apj] {10.3847/1538-4357/aba1e6}, \href
  {https://ui.adsabs.harvard.edu/abs/2020ApJ...899...27Z} {899, 27}

\bibitem[\protect\citeauthoryear{{Zhou} \& {Bayliss}}{{Zhou} \&
  {Bayliss}}{2012}]{2012MNRAS.426.2483Z}
{Zhou} G.,  {Bayliss} D.~D.~R.,  2012, \mn@doi [\mnras]
  {10.1111/j.1365-2966.2012.21817.x}, \href
  {https://ui.adsabs.harvard.edu/abs/2012MNRAS.426.2483Z} {426, 2483}

\bibitem[\protect\citeauthoryear{{von Essen}, {Mallonn}, {Welbanks},
  {Madhusudhan}, {Pinhas}, {Bouy}  \& {Weis Hansen}}{{von Essen}
  et~al.}{2019}]{2019A&A...622A..71V}
{von Essen} C.,  {Mallonn} M.,  {Welbanks} L.,  {Madhusudhan} N.,  {Pinhas} A.,
   {Bouy} H.,   {Weis Hansen} P.,  2019, \mn@doi [\aap]
  {10.1051/0004-6361/201833837}, \href
  {https://ui.adsabs.harvard.edu/abs/2019A&A...622A..71V} {622, A71}

\makeatother
\end{thebibliography}




\appendix

\section{Posterior distributions for the white light curve fits}

\begin{figure*}
\centering
\includegraphics[width=\textwidth,height=0.9\textheight,keepaspectratio]{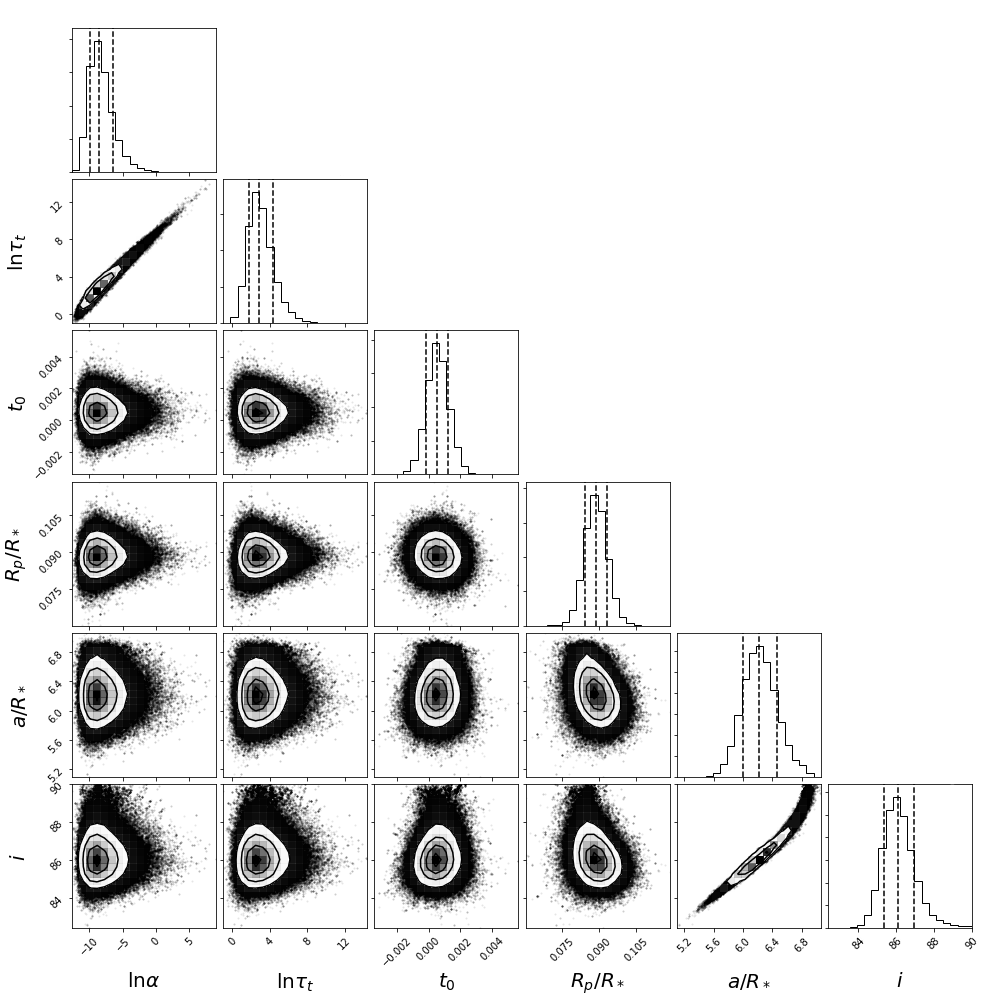}
\caption{The posterior distribution from the first fit of the blue data set.}
\label{fig:blue_corner}
\end{figure*}

\begin{figure*}
\centering
\includegraphics[width=\textwidth,height=0.9\textheight,keepaspectratio]{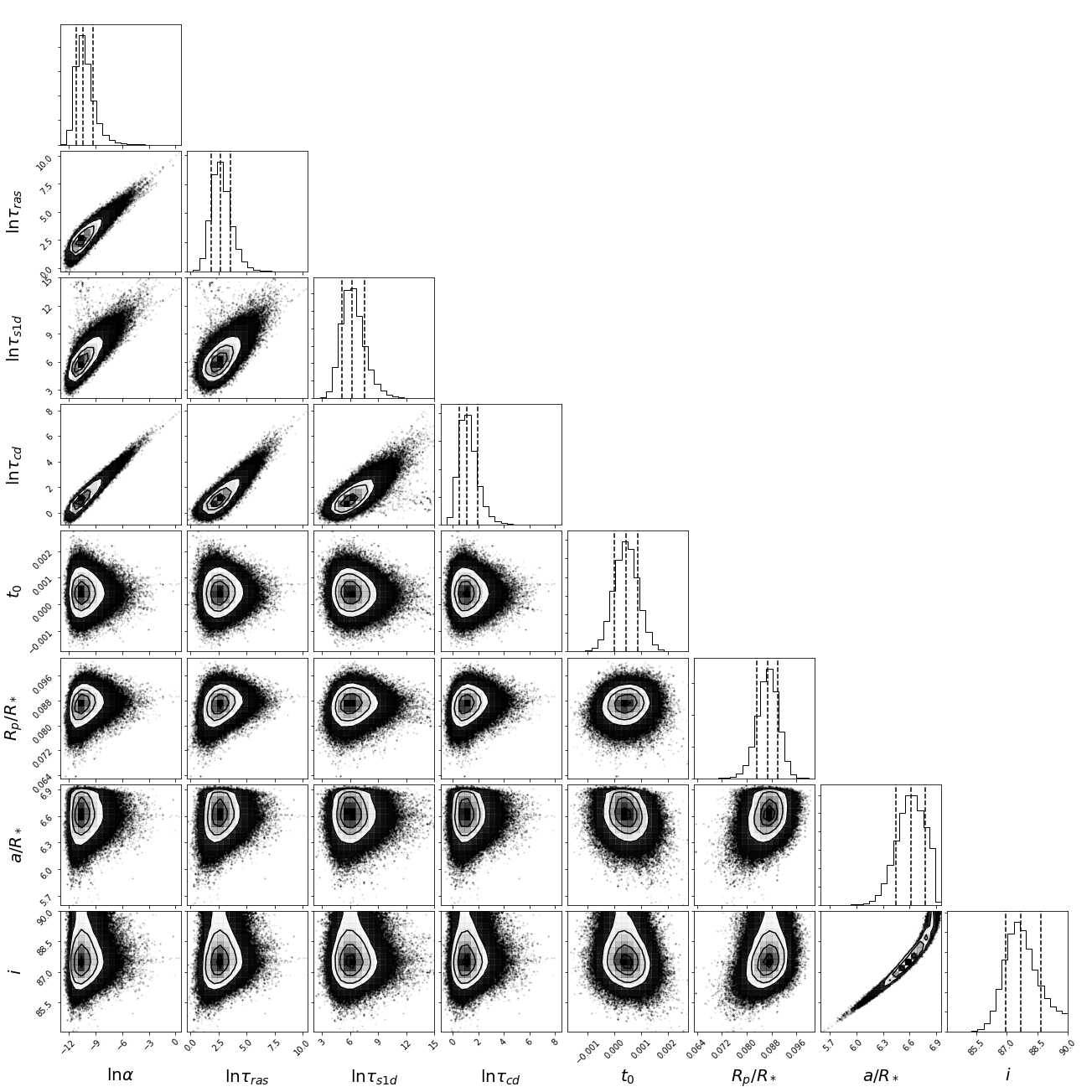}
\caption{The posterior distribution from the first fit of the red data set.}
\label{fig:red_corner}
\end{figure*}

\section{Auxiliary variables}

\begin{figure*}
\centering
\includegraphics[width=\textwidth,height=0.9\textheight,keepaspectratio]{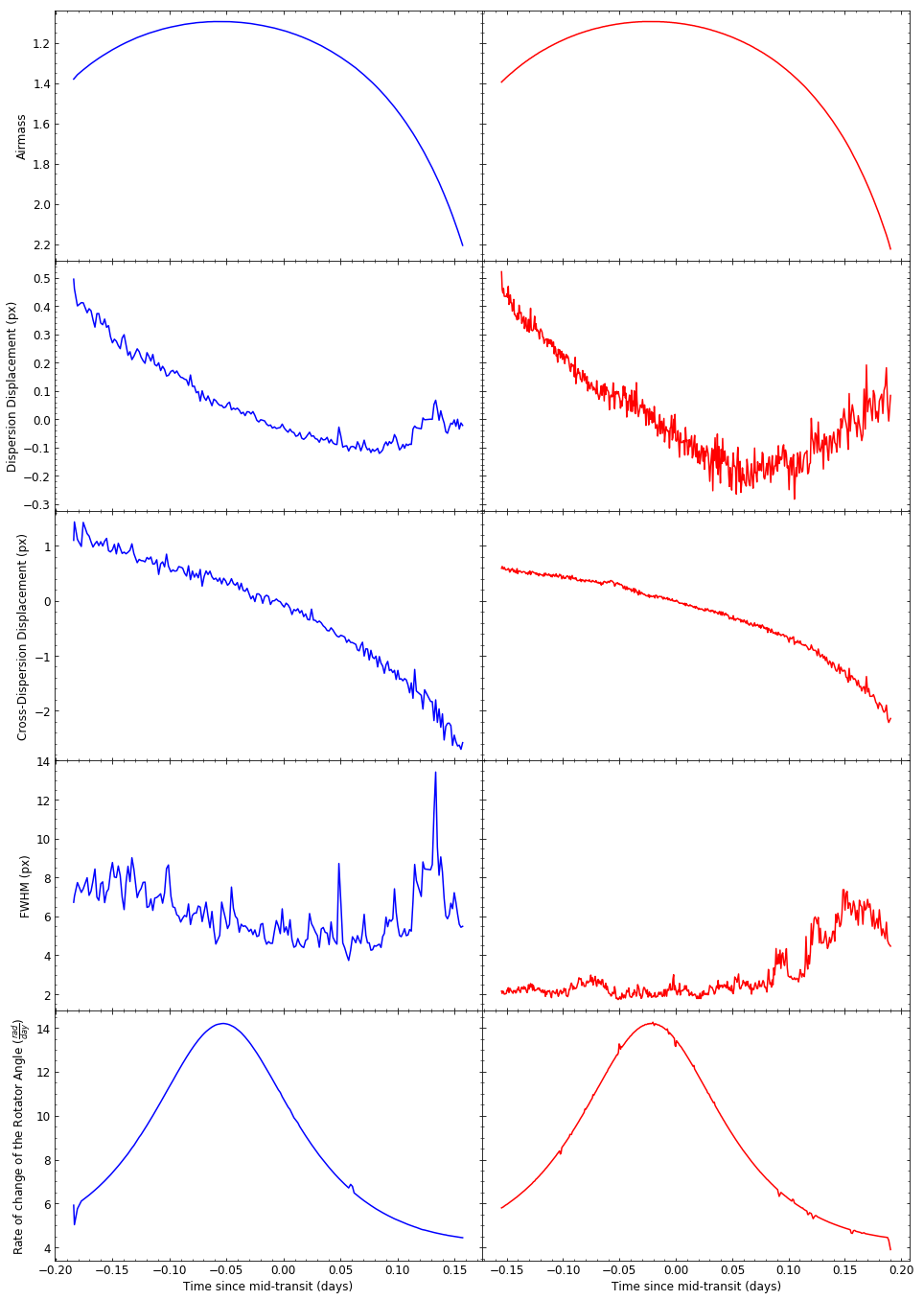}
\caption{Detrending variables as a function of time for the blue (left) and red (right) data sets. From top to bottom, the figure includes airmass, displacements in the x and y axis, FWHM, and the speed of the rotator angle.}
\label{fig:auxiliary_variables}
\end{figure*}

\section{Transmission spectrum parameters}

\begin{table}
\centering
\caption{Transmission spectrum and limb darkening coefficients from the combined spectroscopic light curves.}
\label{tab:wasp88_spectroscopic}
\begin{tabular}{lccc}
\hline
\hline
Wavelength Range (\AA) & $R_\mathrm{p}/R_*$ & $u_1$ & $u_2$\\
\hline
$4413-4653$ & $0.0916^{+0.0041}_{-0.0046}$ & $0.392^{+0.078}_{-0.085}$ & 0.336\\[2pt]
$4653-4733$ & $0.0939^{+0.0082}_{-0.0040}$ & $0.422^{+0.060}_{-0.059}$ & 0.355\\[2pt]
$4733-4813$ & $0.1009^{+0.0039}_{-0.0079}$ & $0.431^{+0.069}_{-0.075}$ & 0.347\\[2pt]
$4813-4893$ & $0.0885^{+0.0044}_{-0.0039}$ & $0.264^{+0.073}_{-0.081}$ & 0.416\\[2pt]
$4893-4973$ & $0.0880^{+0.0038}_{-0.0040}$ & $0.348^{+0.077}_{-0.076}$ & 0.374\\[2pt]
$4973-5053$ & $0.0873^{+0.0035}_{-0.0038}$ & $0.449^{+0.063}_{-0.066}$ & 0.340\\[2pt]
$5053-5133$ & $0.0873^{+0.0036}_{-0.0042}$ & $0.281^{+0.075}_{-0.093}$ & 0.358\\[2pt]
$5133-5213$ & $0.0855^{+0.0040}_{-0.0047}$ & $0.333^{+0.087}_{-0.095}$ & 0.353\\[2pt]
$5213-5293$ & $0.0904^{+0.0040}_{-0.0045}$ & $0.336^{+0.078}_{-0.084}$ & 0.371\\[2pt]
$5293-5373$ & $0.0886^{+0.0028}_{-0.0029}$ & $0.255^{+0.062}_{-0.068}$ & 0.365\\[2pt]
$5373-5453$ & $0.0892^{+0.0028}_{-0.0029}$ & $0.170^{+0.069}_{-0.074}$ & 0.400\\[2pt]
$5453-5533$ & $0.0893^{+0.0028}_{-0.0028}$ & $0.268^{+0.057}_{-0.060}$ & 0.361\\[2pt]
$5533-5613$ & $0.0904^{+0.0021}_{-0.0022}$ & $0.260^{+0.048}_{-0.052}$ & 0.368\\[2pt]
$5613-5693$ & $0.0890^{+0.0025}_{-0.0027}$ & $0.183^{+0.058}_{-0.066}$ & 0.369\\[2pt]
$5693-5773$ & $0.0847^{+0.0029}_{-0.0031}$ & $0.203^{+0.064}_{-0.068}$ & 0.376\\[2pt]
$5773-5853$ & $0.0878^{+0.0023}_{-0.0024}$ & $0.242^{+0.050}_{-0.052}$ & 0.358\\[2pt]
$5853-5933$ & $0.0867^{+0.0023}_{-0.0025}$ & $0.247^{+0.054}_{-0.058}$ & 0.380\\[2pt]
$5933-6013$ & $0.0861^{+0.0022}_{-0.0023}$ & $0.198^{+0.055}_{-0.061}$ & 0.377\\[2pt]
$6013-6093$ & $0.0874^{+0.0024}_{-0.0025}$ & $0.126^{+0.061}_{-0.066}$ & 0.374\\[2pt]
$6093-6173$ & $0.0864^{+0.0025}_{-0.0024}$ & $0.165^{+0.058}_{-0.064}$ & 0.358\\[2pt]
$6173-6253$ & $0.0874^{+0.0028}_{-0.0027}$ & $0.205^{+0.065}_{-0.069}$ & 0.378\\[2pt]
$6253-6333$ & $0.0914^{+0.0056}_{-0.0053}$ & $0.256^{+0.087}_{-0.095}$ & 0.364\\[2pt]
$6333-6413$ & $0.0862^{+0.0036}_{-0.0034}$ & $0.070^{+0.091}_{-0.108}$ & 0.375\\[2pt]
$6413-6493$ & $0.0875^{+0.0029}_{-0.0026}$ & $0.299^{+0.062}_{-0.061}$ & 0.379\\[2pt]
$6493-6573$ & $0.0887^{+0.0066}_{-0.0047}$ & $-0.074^{+0.143}_{-0.137}$ & 0.391\\[2pt]
$6573-6653$ & $0.0866^{+0.0034}_{-0.0030}$ & $0.222^{+0.067}_{-0.074}$ & 0.390\\[2pt]
$6653-6733$ & $0.0850^{+0.0031}_{-0.0029}$ & $0.207^{+0.069}_{-0.075}$ & 0.369\\[2pt]
$6733-6813$ & $0.0873^{+0.0034}_{-0.0029}$ & $0.170^{+0.071}_{-0.077}$ & 0.361\\[2pt]
$6813-6973$ & $0.0873^{+0.0028}_{-0.0026}$ & $0.267^{+0.058}_{-0.058}$ & 0.363\\[2pt]
$6973-7053$ & $0.0867^{+0.0033}_{-0.0029}$ & $0.236^{+0.071}_{-0.071}$ & 0.363\\[2pt]
$7053-7133$ & $0.0862^{+0.0033}_{-0.0031}$ & $0.176^{+0.077}_{-0.082}$ & 0.361\\[2pt]
$7133-7213$ & $0.0844^{+0.0029}_{-0.0030}$ & $0.172^{+0.074}_{-0.083}$ & 0.351\\[2pt]
$7213-7293$ & $0.0852^{+0.0029}_{-0.0030}$ & $0.131^{+0.077}_{-0.091}$ & 0.360\\[2pt]
$7293-7373$ & $0.0835^{+0.0030}_{-0.0031}$ & $0.131^{+0.081}_{-0.092}$ & 0.361\\[2pt]
$7373-7453$ & $0.0864^{+0.0030}_{-0.0029}$ & $0.161^{+0.070}_{-0.078}$ & 0.360\\[2pt]
$7453-7533$ & $0.0843^{+0.0030}_{-0.0032}$ & $0.212^{+0.080}_{-0.086}$ & 0.357\\[2pt]
$7533-7693$ & $0.0852^{+0.0028}_{-0.0027}$ & $0.150^{+0.066}_{-0.071}$ & 0.361\\[2pt]
$7693-7773$ & $0.0863^{+0.0039}_{-0.0038}$ & $0.178^{+0.086}_{-0.099}$ & 0.351\\[2pt]
$7773-7853$ & $0.0856^{+0.0047}_{-0.0051}$ & $0.230^{+0.102}_{-0.116}$ & 0.362\\[2pt]
$7853-7933$ & $0.0877^{+0.0045}_{-0.0048}$ & $0.383^{+0.084}_{-0.087}$ & 0.355\\[2pt]
$7933-8013$ & $0.0854^{+0.0046}_{-0.0044}$ & $0.241^{+0.101}_{-0.109}$ & 0.352\\[2pt]
$8013-8093$ & $0.0828^{+0.0046}_{-0.0047}$ & $0.396^{+0.080}_{-0.088}$ & 0.348\\[2pt]
$8093-8173$ & $0.0875^{+0.0043}_{-0.0041}$ & $0.257^{+0.086}_{-0.096}$ & 0.353\\[2pt]
$8173-8253$ & $0.0848^{+0.0018}_{-0.0019}$ & $0.182^{+0.069}_{-0.059}$ & 0.349\\[2pt]
$8253-8333$ & $0.0792^{+0.0030}_{-0.0034}$ & $0.224^{+0.084}_{-0.097}$ & 0.349\\[2pt]
\hline
\end{tabular}
\end{table}


\bsp	
\label{lastpage}
\end{document}